\documentclass[a4paper, fleqn]{scrartcl}

\usepackage{amsmath}
\usepackage{amssymb}
\usepackage{graphicx}
\usepackage{setspace}
\usepackage[sort&compress]{natbib}

\usepackage[T1]{fontenc}
\usepackage{mathptmx}
\usepackage[scaled=0.90]{helvet}
\usepackage{courier}

\usepackage[sort&compress]{natbib}
  \bibpunct{[}{]}{,}{n}{}{;}
\usepackage[amssymb, mediumspace]{SIunits}

\usepackage{color}
  \definecolor{myDarkRed}{rgb}{0.5,0,0}       
  \definecolor{myDarkBlue}{rgb}{0,0,0.6}      
  \definecolor{myDarkGreen}{rgb}{0,0.4,0}     
  \definecolor{myDarkGray}{rgb}{0.4,0.4,0.4}  
  \definecolor{black}{rgb}{0,0,0}
  \definecolor{white}{rgb}{1,1,1}

\usepackage[%
  colorlinks={true},%
  linkcolor={myDarkBlue},%
  citecolor={myDarkGreen},%
  urlcolor={myDarkBlue},%
  pdfauthor={Matthias H\"olzl},%
  pdflang={en-US},%
  baseurl={},%
  breaklinks={true},%
  hyperfootnotes={false}%
  ]{hyperref}

\usepackage{doi}

\usepackage{lineno}
\linenumbers

\begin{document}

\setlength{\parindent}{0em}
\setlength{\parskip}{4ex plus2ex minus1ex}

\newcommand{\pderiv}[2]{\frac{\partial #1}{\partial #2}} 
\newcommand{\dderiv}[2]{\frac{d #1}{d #2}}               

\begin{center}
\textbf{\huge Reduced-MHD Simulations of Toroidally and Poloidally Localized ELMs}

{\Large M.~H\"olzl$^1$, S.~G\"unter$^1$, R.P.~Wenninger$^2$, W.-C.~M\"uller$^1$, G.T.A.~Huysmans$^3$, K.~Lackner$^1$, I.~Krebs$^1$, and the ASDEX Upgrade Team$^1$}

\textit{
$^1$Max-Planck-Institut f\"ur Plasmaphysik, EURATOM Association, Boltzmannstra\ss e 2, 85748 Garching, Germany \\
$^2$Universit\"atssternwarte der Ludwig-Maximilians-Universit\"at, 81679 M\"unchen, Germany \\
$^3$ITER Organisation, Route de Vinon sur Verdon, St-Paul-lez-Durance, France
}
\end{center}

\section*{Abstract}

We use the non-linear reduced-MHD code JOREK to study ELMs in the geometry of the
ASDEX Upgrade tokamak. Toroidal mode numbers, poloidal filament sizes, and radial
propagation speeds of filaments into the scrape-off layer are in good agreement with
observations for type-I ELMs in ASDEX Upgrade. The observed instabilities exhibit
a toroidal and poloidal localization of perturbations which is compatible with the
``solitary magnetic perturbations'' recently discovered in ASDEX Upgrade
[R.Wenninger et.al., Solitary Magnetic Perturbations at the ELM Onset, Nucl.Fusion,
accepted, preprint at \url{http://arxiv.org/abs/1202.3603}]. This localization
can only be described in numerical simulations with high toroidal resolution.

\newpage

\section{Introduction}\label{:intro}

Edge localized modes~(ELMs) accompany the high-confinement mode~(H-mode)
in tokamak fusion plasmas. As they could cause a potentially destructive
heat-load on divertor plates and wall structures in large fusion
devices~\citep{Loarte2003,Klimov2011}, detailed understanding of these instabilities
and of mitigation-techniques is important for a successful operation of ITER. 
Non-linear MHD-simulations with the JOREK code~\citep{Huysmans2007,Huysmans2008,
Huysmans2009,Pamela2010,Huysmans2010,Pamela2011,Hoelzl2011}, which is also used
for the present article, and other codes like BOUT++~\citep{Dudson2011},
NIMROD~\citep{Pankin2007}, or M3D~\citep{Sugiyama2010} can make an important contribution
after successful benchmarks with measurements in existing tokamaks.

In the present article, a comparison between simulations with the non-linear
finite-element code JOREK~\citep{Czarny2008} and observations in the ASDEX Upgrade
tokamak~\citep{HERMANN03A} is started. We concentrate
on the early phase of ELMs. JOREK solves
the reduced MHD equations in realistic X-point geometries as described
in Section~\ref{:jorek}. ASDEX Upgrade is
equipped with a unique set of edge diagnostics that allows to
investigate ELM crashes with high spatial and temporal resolutions~\citep{Wolfrum2011}.
This provides excellent possibilities for theory-experiment comparisons.
Emphasis is put on simulations with high toroidal resolution (many toroidal
modes at toroidal periodicity 1) to treat the coupling
between various toroidal modes properly. This way, aspects can be identified
that are described well already at low toroidal resolution (few toroidal modes
at a high toroidal periodicity) while others
are influenced significantly by the non-linear toroidal mode-coupling.

The article is structured as follows. Section~\ref{:jorek}
describes the non-linear MHD-code JOREK. Physical
parameters and technical details of the numerical simulations are given
in Section~\ref{:simulations}. Our observations and findings made in the
simulated instabilities are presented
in Section~\ref{:results}. Subsequently, Section~\ref{:comparison} describes how
these results compare to experimental measurements.
Finally, Section~\ref{:conclusions} summarizes and gives a brief outlook.

\section{JOREK Code}\label{:jorek}

The simulations are carried out with the single-fluid reduced-MHD
model of the JOREK code.
Section~\ref{:jorek:equations} describes the equations
solved in this model. For more details on the derivation,
refer to Reference~\citep{Strauss1997} and Appendix~\ref{:app:inductioneq}.
Spatial and temporal discretizations are
briefly addressed in Section~\ref{:jorek:discretization}.

\subsection{Reduced-MHD Equations}\label{:jorek:equations}

\begin{table}
\caption{The normalization of quantities in JOREK is listed, it
corresponds to choosing scale factors $B_0=\unit{1}{\tesla}$ and $R_0=\unit{1}{\metre}$.
Variable names with subscript ``SI'' denote quantities in SI units, while
variables without this subscript are the ones used in JOREK. In the presented
simulations, $n_0=\unit{6\cdot10^{19}}{\metre^{-3}}$ and
$\rho_0=\unit{2\cdot10^{-7}}{\kilo\gram\usk\metre^{-3}}$. The
magnetic constant is denoted $\mu_0$ and the Boltzmann constant $k_B$.}
\label{table:norm}
\begin{tabular}{l l l}
\hline
$R_\text{SI}~[\metre]$ & $=R$ & Major radius \\
$Z_\text{SI}~[\metre]$ & $=Z$ & Vertical coordinate \\
$\mathbf{B}_\text{SI}~[\tesla]$ & $=\mathbf{B}$ & Magnetic field vector;
  see Eq.~\eqref{eq:B-def} \\
$\Psi_\text{SI}~[\tesla\usk\square\metre]$ & $=\Psi$ & Poloidal magnetic flux \\
$j_{\phi, \text{SI}}~[\ampere\usk\metre^{-2}]$ & $=-j/(R\;\mu_0)$ & Toroidal current density;
  $j_{\phi, \text{SI}}=\mathbf{j}_{\text{SI}}\cdot\hat{\mathbf{e}}_\phi$ \\
$n_\text{SI}~[\metre^{-3}]$ & $=\rho\;n_0$ & Particle density \\
$\rho_\text{SI}~[\kilo\gram\usk\metre^{-3}]$ & $=\rho\;\rho_0$ & Mass density = ion mass $\times$ particle density \\
$T_\text{SI}~[\kelvin]$ & $=T/(k_B\;\mu_0\;n_0)$ & Temperature = electron + ion temperature \\
$p_\text{SI}~[\newton\usk\metre^{-2}]$ & $=\rho\;T/\mu_0$ & Plasma pressure \\
$\mathbf{v}_\text{SI}~[\meter\usk\second^{-1}]$ & $=\mathbf{v}/\sqrt{\mu_0\rho_0}$ & Velocity vector;
  see Eq.~\eqref{eq:v-def} \\
$v_{||, \text{SI}}~[\meter\usk\second^{-1}]$ & $=v_{||}\cdot B_\text{SI}/\sqrt{\mu_0\rho_0}$ & Parallel velocity component,
  where $B_\text{SI}=|\mathbf{B}_\text{SI}|$ \\
$u_\text{SI}~[\meter\usk\second^{-1}]$ & $=u/\sqrt{\mu_0\rho_0}$ & Velocity stream function \\
$\omega_{\phi, \text{SI}}~[\meter^{-1}\usk\second^{-1}]$ & $=\omega/\sqrt{\mu_0\rho_0}$ & Toroidal vorticity;
  see Eq.~\eqref{eq:omegadef} \\
$t_\text{SI}~[\second]$ & $=t\cdot\sqrt{\mu_0\rho_0}$ & Time \\
$\gamma_\text{SI}~[\second^{-1}]$ & $=\gamma/\sqrt{\mu_0\rho_0}$ & Growth rate;
  $\gamma_\text{SI}=\ln[E_\text{SI}(t_2)/E_\text{SI}(t_1)]/[2\Delta t_\text{SI}]$; Energy $E_\text{SI} [\joule]$ \\
$\eta_\text{SI}~[\ohm\usk\meter]$ & $=\eta\cdot\sqrt{\mu_0/\rho_0}$ & Resistivity \\
$\nu_\text{SI}~[\kilo\gram\usk\metre^{-1}\usk\second^{-1}]$ & $=\nu\cdot\sqrt{\rho_0/\mu_0}$ & Dynamic viscosity \\
$D_\text{SI}~[\square\metre\usk\second^{-1}]$ & $=D/\sqrt{\mu_0\rho_0}$ & Particle diffusivity ($||$ or $\bot$) \\
$K_\text{SI}~[\metre^{-1}\usk\second^{-1}]$ & $=K\cdot n_0/\sqrt{\mu_0\rho_0}$ & Heat diffusivity ($||$ or $\bot$),
  where $\chi_\text{SI}~[\square\metre\usk\second^{-1}]=K_\text{SI}/n_\text{SI}$ \\
$S_{T, \text{SI}}~[\watt\usk\metre^{-3}]$ & $=S_T/\sqrt{\mu_0^3\rho_0}$ & Heat source \\
$S_{\rho, \text{SI}}~[\kilo\gram\usk\second^{-1}\usk\metre^{-3}]$ & $=S_\rho\cdot\sqrt{\rho_0/\mu_0}$ & Particle source \\
\hline
\end{tabular}
\end{table}

Seven physical variables are treated:
poloidal flux $\Psi$, stream function $u$, toroidal current density $j$,
toroidal vorticity $\omega$, density $\rho$,
temperature $T$, and velocity $v_{||}$ along magnetic field lines. The
normalization of the relevant quantities is listed in
Table~\ref{table:norm}.

Variables $j$ and $\omega$ are connected to $\Psi$ and $u$ by the definition equations

\begin{align}
  j &= \Delta^*\Psi = R^2\;\nabla_{pol}\cdot\left(R^{-2}\;\nabla_{pol}\Psi\right)
     = R \dderiv{}{R} \left( \frac{1}{R} \dderiv{\Psi}{R} \right) + \frac{d^2\Psi}{dZ^2}, \label{eq:jdef} \\
  \omega &= \nabla_{pol}^2~u = \frac{1}{R}\dderiv{}{R}\left(R\dderiv{u}{R}\right) + \frac{d^2u}{dZ^2} \label{eq:omegadef},
\end{align}

where $\nabla_\text{pol}$ denotes the del-operator in the poloidal plane,
$R$ the major radius, and $Z$ the vertical coordinate.
The time-evolution of the remaining five free variables is described by the following
set of equations (called physics-model ``302'' in JOREK):

\begin{align}
      \pderiv{\Psi}{t}&=\eta j
        -R[u,\Psi]-F_0\pderiv{u}{\phi}, \label{eq:psievol} \\
      \pderiv{\rho}{t}&=-\nabla\cdot(\rho\mathbf{v})+\nabla\cdot(D_\bot\nabla_\bot\;\rho)+S_\rho, \\
      \rho\pderiv{T}{t}&=-\rho\mathbf{v}\cdot\nabla T-(\kappa-1)p\nabla\cdot\mathbf{v}
        +\nabla\cdot\left(K_\bot\nabla_\bot\;T+K_{||}\nabla_{||}T\right)+S_T, \\
      \hat{\mathbf{e}}_\phi\cdot\nabla\times\bigg\{
        \rho\pderiv{\mathbf{v}}{t}&=
          -\rho(\mathbf{v}\cdot\nabla)\mathbf{v}-\nabla p
          +\mathbf{j}\times\mathbf{B}+\nu\Delta\mathbf{v}
      \bigg\}, \\
      \mathbf{B}\cdot\bigg\{\rho\pderiv{\mathbf{v}}{t}&=
        -\rho(\mathbf{v}\cdot\nabla)\mathbf{v}-\nabla p
          +\mathbf{j}\times\mathbf{B}+\nu\Delta\mathbf{v}
      \bigg\}. \label{eq:vparevol}
\end{align}
In every time-step, Equations~(\ref{eq:jdef}--\ref{eq:vparevol}) are solved simultaneously in weak form
as a large sparse implicit system of equations.
The velocity vector is defined as

\begin{equation}\label{eq:v-def}
  \mathbf{v}=-R\nabla u \times\hat{\mathbf{e}}_\phi+v_{||}\;\mathbf{B},
\end{equation}

the magnetic field vector as

\begin{equation}\label{eq:B-def}
  \mathbf{B}=(F_0\;\hat{\mathbf{e}}_\phi+\nabla\Psi\times\hat{\mathbf{e}}_\phi)/R,
\end{equation}

the pressure is $p = \rho T$, and $\kappa=5/3$ denotes the ratio of specific heats. Here,
$\hat{\mathbf{e}}_\phi$ denotes the normalized toroidal basis vector. The toroidal
magnetic field $B_\phi=F_0/R$ is fixed and cannot change with time. The poloidal velocity, i.e.,
the velocity vector in the poloidal plane, is denoted $\mathbf{v}_\text{pol}=$.
The parallel gradient is given by $\nabla_{||}=\mathbf{b}(\mathbf{b}\cdot\nabla)$,
where $\mathbf{b}=\mathbf{B}/|\mathbf{B}|$, and the perpendicular gradient
by $\nabla_\bot=\nabla-\nabla_{||}$. The Poisson bracket $[u,\Psi]$ is defined as
$\pderiv{u}{R}\pderiv{\Psi}{Z}-\pderiv{u}{Z}\pderiv{\Psi}{R}$. Note, that the poloidal
components of the velocity in this set of equations are determined only by the
$\mathbf{E}\times\mathbf{B}$-drift term. As a result, $u$ acts as a velocity stream
function and (except for a factor $F_0$) also as electric potential.

Ideal-wall boundary conditions are implemented where the boundary of the computational domain
is parallel to the magnetic flux surfaces. At the divertor targets, where the flux surfaces
intersect the computational boundary, modified
Bohm boundary conditions apply~\citep{Stangeby1995,Huysmans2008}.

\subsection{Discretization}\label{:jorek:discretization}

\begin{table}
\caption{The toroidal mode numbers resolved in simulations with different periodicities are listed.}
\label{table:ntor}
\begin{tabular}{c c l}
  \hline
  Periodicity & Resolved $n$-modes \\
  \hline
  8           & $0,8,16$ \\
  4           & $0,4,8,12,16$ \\
  2           & $0,2,4,\dots,12,14,16$ \\
  1           & $0,1,2,\dots,14,15,16$ \\
  \hline
\end{tabular}
\end{table}

The poloidal plane is discretized by 2D Bezier finite elements with four degrees
of freedom per grid node and physical variable~\citep{Czarny2008}, while a Fourier
decomposition is applied toroidally. The number of toroidal modes
resolved in the simulations and the assumed toroidal periodicity of the system can
be chosen separately. A periodicity equal to one means that the solution is computed
for the whole torus. For larger periodicities, only a toroidal section
of the torus is resolved. The modes included in the presented
simulations are listed in Table~\ref{table:ntor}.

The temporal discretization is performed by a fully implicit second-order
linearized Crank-Nicholson scheme~\citep{Hirsch1989}. In the resulting large
sparse system, all physical equations and all toroidal harmonics are coupled.
It is solved by an iterative GMRES-method, where a physics-based preconditioning
is applied at the beginning of each GMRES solver step. In the preconditioning,
the coupling between the sub-matrices corresponding to individual toroidal
harmonics is neglected which allows to solve each sub-system separately. This
is performed using the direct solver PaStiX~\citep{PASTIX}.

\section{Simulations}\label{:simulations}

Simulations of edge-localized modes are one of the most challenging
tasks in fusion MHD numerics. The problem must be treated in realistic X-point
geometry as the mode-affected region extends from inside
the H-mode pedestal into the scrape-off layer, the vacuum region, and to the divertor
legs. High spatial resolutions in all dimensions are required due to the small
scales of the structures and the large radial gradients of equilibrium
quantities at the pedestal.
Thus, as a consequence of limited computational resources, not all aspects
of an experiment can be described realistically
in simulations so far. For instance,
simulations with high resolution in radial and poloidal directions, i.e., with
a large number of 2D Bezier finite-elements in the case of JOREK,
render important investigations at more realistic
plasma resistivities possible (e.g., Ref.~\citep{Pamela2011}), but only
at a very limited number of toroidal Fourier harmonics.

\begin{table}
\caption{Core values for plasma resistivity and viscosity are listed for the
simulations denoted \texttt{eta5} and \texttt{eta6}. Both quantities are modeled with a
$T^{-3/2}$ dependence and are chosen significantly larger than in experiments due to
computational restrictions. In ASDEX Upgrade, the core resistivity is typically about
$\unit{10^{-8}}{\ohm\usk\meter}$.}
\label{table:eta}
\begin{tabular}{c c c c c}
  \hline
  Run              & $\eta_\text{SI}$ [$\unit{}{\ohm\usk\meter}$]                   & $\nu_\text{SI}$ [$\unit{}{\square\meter\per\second}$]       \\
  \hline
  \texttt{eta5}    & $5\times10^{-5}$ & $7.5\times10^{-5}$   \\
  \texttt{eta6}    & $5\times10^{-6}$ & $7.5\times10^{-6}$   \\
  \hline
\end{tabular}
\end{table}

\begin{figure}
\centering
  \includegraphics[width=0.3\textwidth]{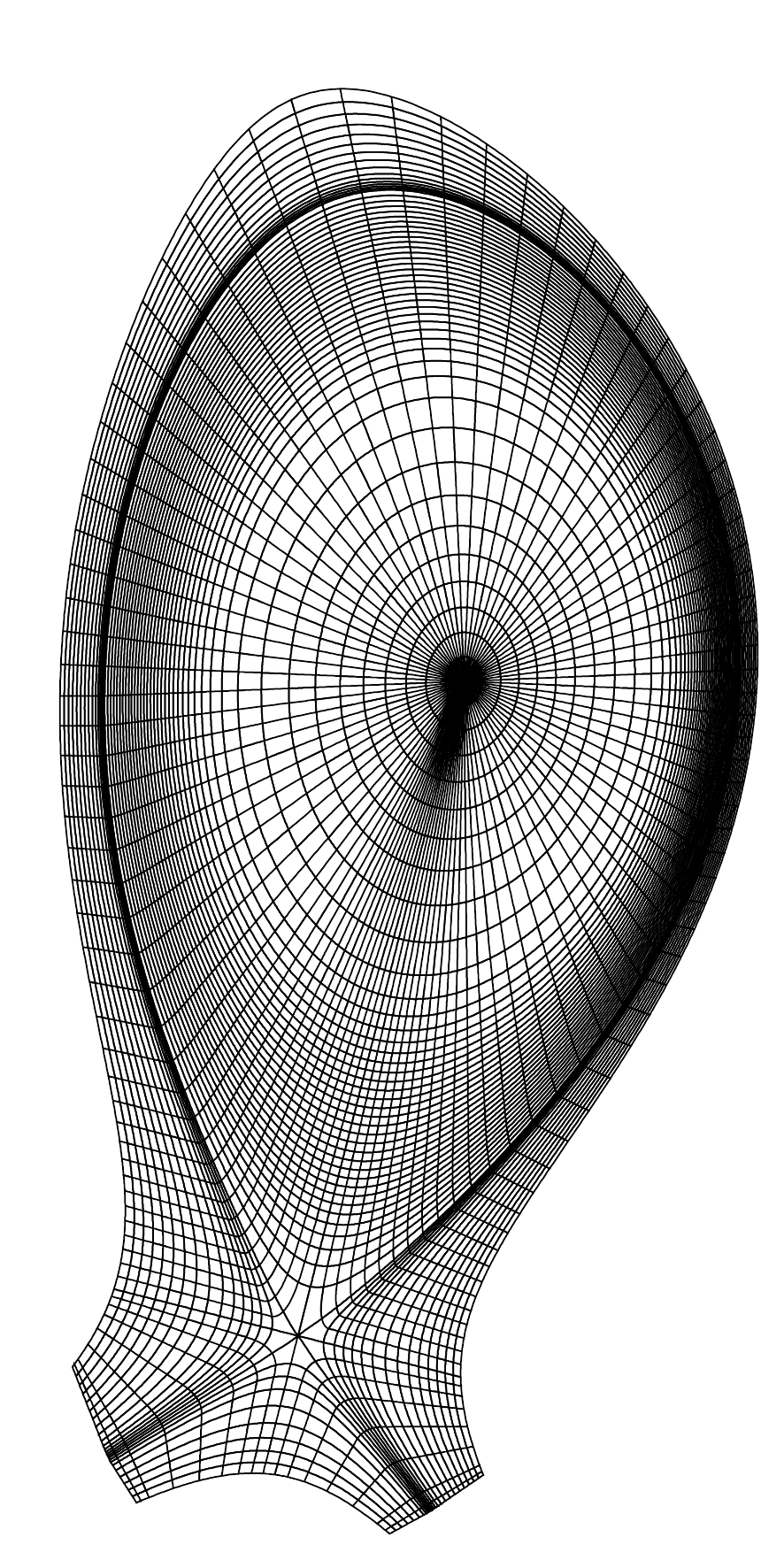}
\caption{The flux-surface aligned X-point grid with $5500$
  Bezier finite-elements is shown. The number of grid points are: $96$ poloidal
  points, $40$ radial points inside the separatrix, $15$ radial points
  outside the separatrix, $9$ ``radial'' points in the private flux
  region, and $9$ grid points along the divertor legs. For the \texttt{eta6}
  simulations, these numbers are all increased by a factor of $\sqrt{2}$, leading
  to about $11000$ Bezier elements.}
\label{fig:grid}
\end{figure}

For this work, a different choice was made: The focus is put on high toroidal
resolution. This is done to investigate the influence of toroidal
mode-coupling onto the non-linear evolution of an ELM.
The mode numbers resolved in the simulations are listed in
Table~\ref{table:ntor}. All runs resolve the $n=0,\dots,16$ range but with
different periodicities. The relatively high number of toroidal modes involved
limits the possible radial and poloidal resolutions:
Most simulations are carried out with about $5500$ Bezier
elements. The corresponding finite-element
grid is shown in  Figure~\ref{fig:grid}. Only for the simulations with
lower plasma resistivity (denoted ``\texttt{eta6}'' runs,
see next paragraph for details), the number of
Bezier elements is increased by a factor of two. Grid accumulation is
used to increase the resolution radially around the separatrix and poloidally
around the X-point.

Due to the comparably low poloidal resolution, only
plasma resistivities significantly larger than in the experiment can be resolved. The
respective simulation parameters are listed in Table~\ref{table:eta}. 
The limited poloidal resolution also reduces the growth rate of modes with high
mode numbers artificially. Thus when increasing the poloidal resolution, the most unstable
mode number would shift towards larger $n$. On the other hand, diamagnetic stabilization
is not taken into account in the simulations. Including this effect would have
a stabilizing effect onto high poloidal mode numbers. The electron diamagnetic frequency
%
%
for $n=10$ is about $\unit{10^5}{\second^{-1}}$ (calculated at a normalized
poloidal flux of $\Psi_N=0.9$).
This is comparable to the fastest linear growth
rates in the simulations (see Section~\ref{:results}). Thus, the most
unstable mode numbers would probably be similar in simulations with higher
poloidal resolution and diamagnetic stabilization taken into account.

All simulations are
based on typical ASDEX Upgrade discharge parameters, details are given
in Section~\ref{:simulations:parameters}. The simulations concentrate on the
early phase of an ELM-crash up to the point where filaments start to form.
The computations are carried out mostly on the HPC-FF cluster located
in J\"ulich, Germany. The \texttt{eta5} simulations with periodicity
$1$ and about $5500$ Bezier elements require at least $102$ compute nodes ($8$ cores
and $24$ GB of memory each) due to memory requirements of the solver and take about
ten thousand CPU hours to complete. The \texttt{eta6} computation with $11000$ Bezier
elements is at the limit of what can be investigated with JOREK on this machine.

\subsection{Physical parameters}\label{:simulations:parameters}

A typical ASDEX Upgrade H-mode discharge with type-I ELMs constitutes
the basis of the simulations: Geometry and profiles are taken from discharge 23221 at 4.7
seconds with a plasma current of $\unit{1}{\mega\ampere}$, $\unit{8}{\mega\watt}$ of
neutral beam injection heating and $\unit{1.5}{\mega\watt}$ of electron
cyclotron resonance heating. The equilibrium
reconstruction with the CLISTE
code~\citep{McCarthy1999,McCarthy1999a} takes into account measured
kinetic profiles. Source terms $S_\rho$ and $S_T$ and perpendicular diffusivities
$D_\bot$ and $K_\bot$ are adjusted such, that the equilibrium does not
change significantly with time. The core temperature is
$k_B T_\text{SI} = k_B(T_{e, \text{SI}} + T_{i, \text{SI}}) = \unit{12.4}{\kilo\electronvolt}$. The safety-factor takes
a value of $q(0)=1$ in the plasma core and $q(\Psi_N=0.95)=4.7$ close to the
separatrix where $\Psi_N=(\Psi-\Psi_\text{axis})/(\Psi_\text{separatrix}-\Psi_\text{axis})$
denotes the normalized poloidal flux. A pure deuterium
plasma with a core density of $\unit{6\cdot10^{19}}{\meter^{-3}}$ is assumed.
The heat diffusion anisotropy, $K_{||}/K_\bot$, takes a value of $7\cdot10^6$
at the separatrix. 

The spatial resolution required for the simulation is, amongst others, determined by
the resistive skin depth $\delta_\text{SI}=\sqrt{2\eta_\text{SI}/(\mu_0\gamma_{SI})}$
which is about $\unit{6}{\milli\meter}$ in \texttt{eta6} simulations.
As the spatial resolution possible in the poloidal plane is limited by
computational resources, realistic plasma resistivities with a resistive skin
depth of about $\unit{0.3}{\milli\meter}$ cannot be resolved (resolving spatial
scales smaller than the ion gyro-radius is of course not reasonable anyway in
MHD-simulations).

\begin{figure}
\centering
  \includegraphics[width=0.85\textwidth]{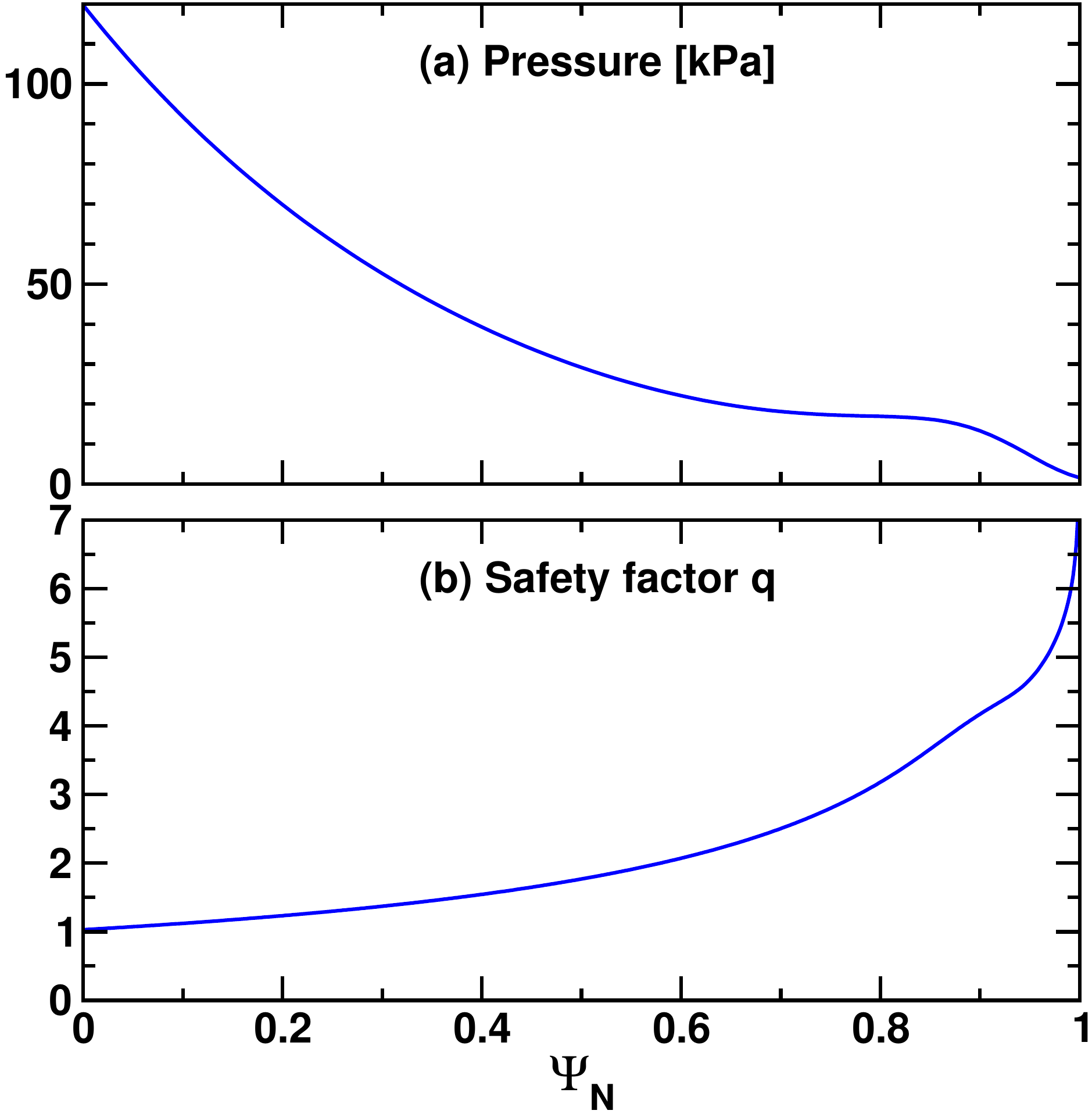}
\caption{Profiles of \textbf{(a)} plasma pressure and \textbf{(b)} safety-factor are shown for the
  plasma equilibrium used in the simulations.}
\label{fig:input-profs}
\end{figure}

The following data are used as inputs for the JOREK simulation:
\begin{itemize}
\item From experimental measurements: Temperature and density profiles, and
toroidal magnetic field strength. The pressure
profile is shown in Figure~\ref{fig:input-profs}a.
\item From CLISTE-equilibrium reconstruction based on experimental
measurements: $F F'$-profile and the values
of the poloidal Flux $\Psi$ at the JOREK computational boundary.
Here, $F=(2\pi/\mu_0)\;J_\text{pol, SI}$ is proportional to the poloidal plasma
current, $J_\text{pol}$, and $F'=dF/d\Psi$. The q-profile of the equilibrium
is shown in Figure~\ref{fig:input-profs}b.
\end{itemize}

In JOREK simulations, the Grad-Shafranov equation is solved first based
on these input parameters. The equilibrium perfectly agrees
with CLISTE (q-profile, flux surfaces, etc.). After that, an ``equilibrium
refinement'' phase is required where the time-evolution equations are solved
only for the $n=0$ mode, with very small time-steps that are gradually increased.
This allows plasma flows to equilibrate~\citep{Pamela2010}. Successively,
the reduced MHD equations are evolved in time, taking into account some or
many toroidal Fourier modes depending on the case considered. Instabilities
then develop out of an initially very small random perturbation.

\section{Simulation Results}\label{:results}

In the following, the simulation results are described and analyzed.
Section~\ref{:results:lowres} addresses simulations with low toroidal
resolution, while Section~\ref{:results:highres} covers the situation at
high toroidal resolution. In the succeeding Section~\ref{:results:loweta},
an attempt towards more realistic plasma resistivities is made.
The simulation results are compared to experimental findings in
Section~\ref{:comparison}. 

\subsection{Low Toroidal Resolution}\label{:results:lowres}

\begin{figure}
\centering
  \includegraphics[width=0.85\textwidth]{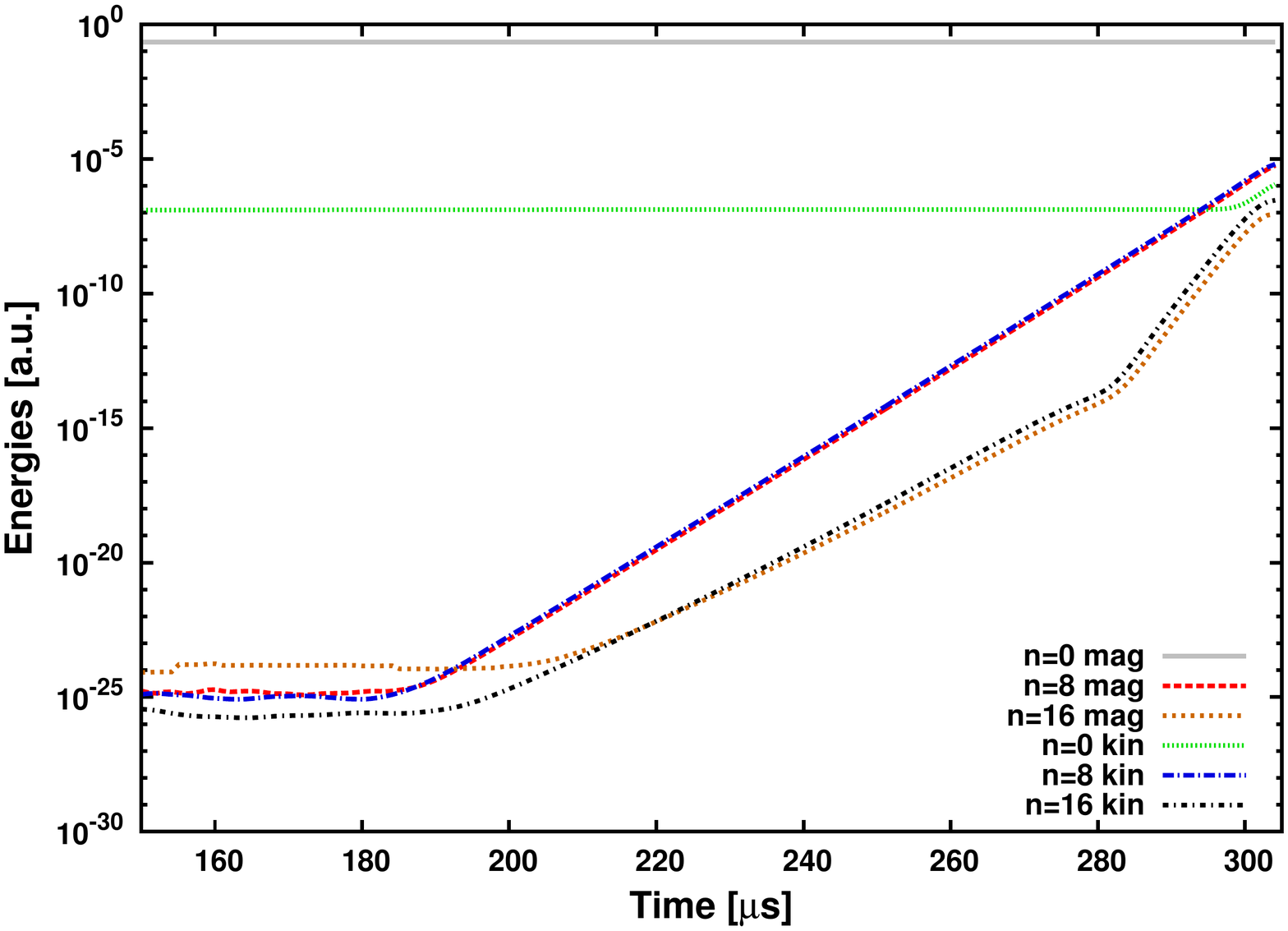}
\caption{Time-traces of the magnetic and kinetic energies contained in the individual toroidal
harmonics are plotted for the simulation with periodicity $8$. 
The $n=8$ mode is linearly more unstable than the
$n=16$ mode and also remains dominant when non-linear saturation sets in.
Due to non-linear mode-interaction, the growth rate of the
$n=16$ mode increases significantly at $t=\unit{284}{\micro\second}$.
The $n=0$ magnetic energy is dominated by the toroidal magnetic field which is
fixed in time as described in Section~\ref{:jorek:equations}.}
\label{fig:energies_ntor5}
\end{figure}

This section provides simulation results for periodicity $8$, where only the
toroidal modes $n=0$, $8$, and $16$ are resolved. A ballooning-like exponentially
growing mode located close to the plasma boundary develops at the
low-field side. As seen in the energy diagnostics shown in
Figure~\ref{fig:energies_ntor5}, the $n=8$ mode is linearly more unstable
(growth rate $\gamma_\text{SI}=\unit{2.0\times10^5}{\second^{-1}}$)
than the $n=16$ mode ($\gamma_\text{SI}\approx\unit{1.5\times10^5}{\second^{-1}}$).
Due to mode-coupling, the structure of the $n=16$
mode changes at $t=\unit{284}{\micro\second}$ in the simulation -- the position
of its maximum amplitude moves radially
from the $q=4$ to the adjacent $q=3.75$ resonant surface.
Hereby, the growth rate of the $n=16$ mode increases significantly to
$\gamma_\text{SI}=\unit{4.3\times10^5}{\second^{-1}}$ which is roughly the double $n=8$
growth rate. The $n=8$ mode also remains dominant at the onset of non-linear mode
saturation ($t\approx\unit{300}{\micro\second}$).

\begin{figure}
\centering
  \includegraphics[height=0.5\textwidth]{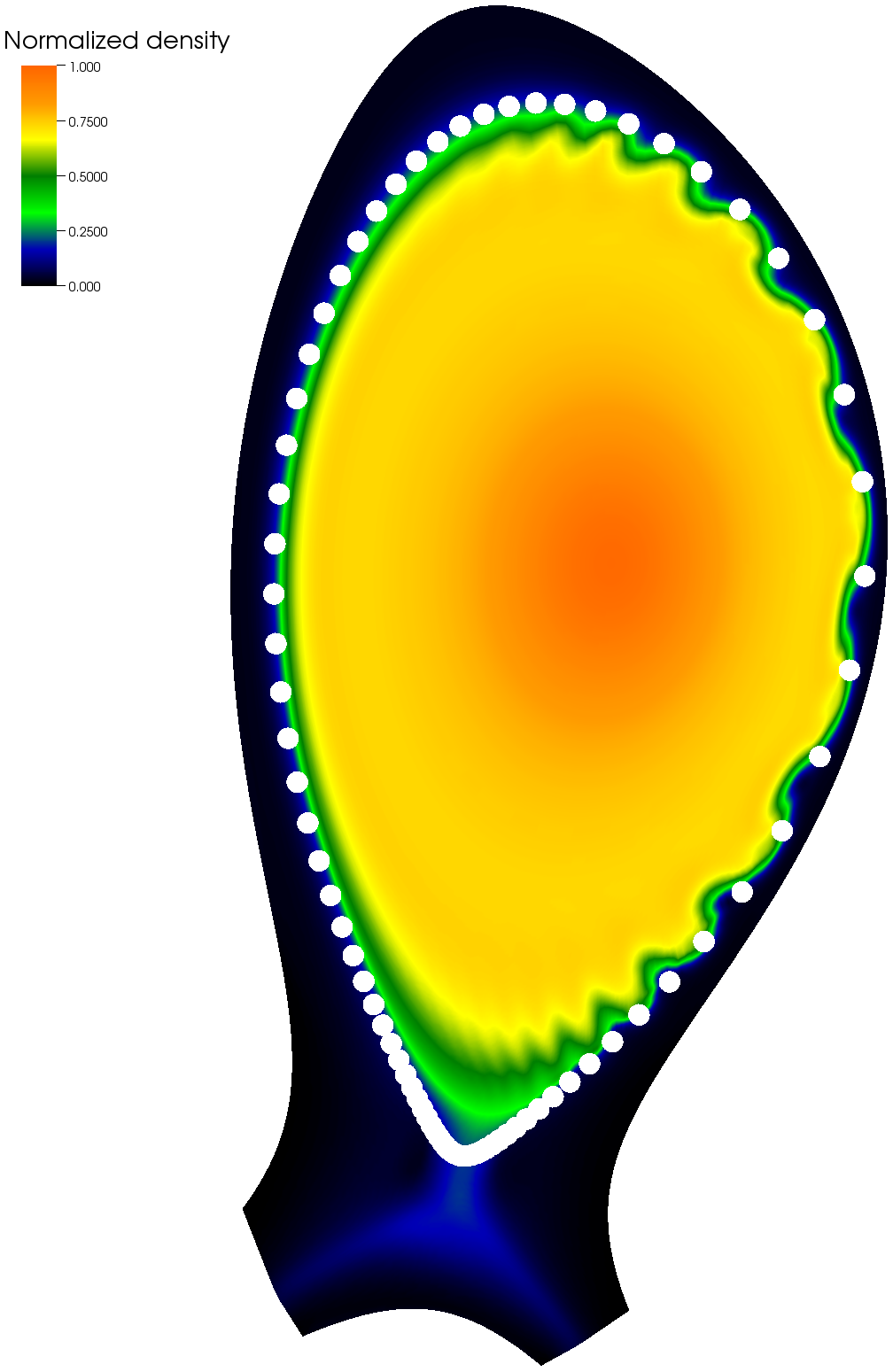}
\caption{The density distribution with developing ballooning-structure in
the simulation with periodicity $8$ is shown at $\unit{298}{\micro\second}$.
Regular ballooning-structures
are observed on the whole low-field side. All ballooning-fingers are roughly
equally wide in poloidal direction in the straight-fieldline
angle $\theta^*$ (the white dots indicate equidistant
distances in $\theta^*$).}
\label{fig:density1}
\end{figure}

The ballooning-structure that develops at the whole low-field side of the
plasma is shown in Figure~\ref{fig:density1} for time point
$\unit{298}{\micro\second}$ in the simulation. The ``density-fingers''
are very regular with a poloidal size of about $\unit{15-20}{\centi\meter}$
at the outer midplane.
The poloidal ``compression'' of the structures in the vicinity of the lower (active)
and the upper (inactive) X-points compared to the outer midplane is a consequence
of field-line stagnation -- the poloidal width of the structures is roughly constant
in the straight-fieldline angle $\theta^*$. In Figure~\ref{fig:density1}, this
can be seen by comparing the density fingers to the white dots which divide the poloidal
circumference into equidistant sections in $\theta^*$.
When the exponentially growing perturbation gets visible in the density
distribution, distortions start to build up which propagate into the
vacuum region as finger-like structures with significantly increased density
due to the $E\times B$ drift. Their radial velocity, measured by
tracing the position at which the density equals 10 percent of the core
density, increases to about $\unit{3}{\kilo\meter\per\second}$ and
saturates at that level.
In the beginning, the density shows sinusoidal
excursions of the density contours which grow
over time (linear phase). As the instability grows
and non-linear saturation sets in (energy growth
rates start to decrease), the density fingers
develop sub-structures. The changing structure
also reflects in a different mode-spectrum, where
the n=16 energies get closer to the n=8 energies
(Figure~\ref{fig:energies_ntor5}).
The ideal-wall boundary conditions contribute to the saturation
of radial velocity when the distance between the mode and the wall
gets significantly smaller than its poloidal wave-length as
mirror-currents build up that slow down the mode-evolution.

\subsection{High Toroidal Resolution}\label{:results:highres}

Now, the same setup as in the previous Section is considered
at periodicity $1$: All toroidal modes
in the range $n=0$ to $16$ are resolved.
The comparison of simulations with different periodicities
allows to identify effects caused by the coupling between toroidal modes.

\begin{figure}
\centering
  \includegraphics[width=0.85\textwidth]{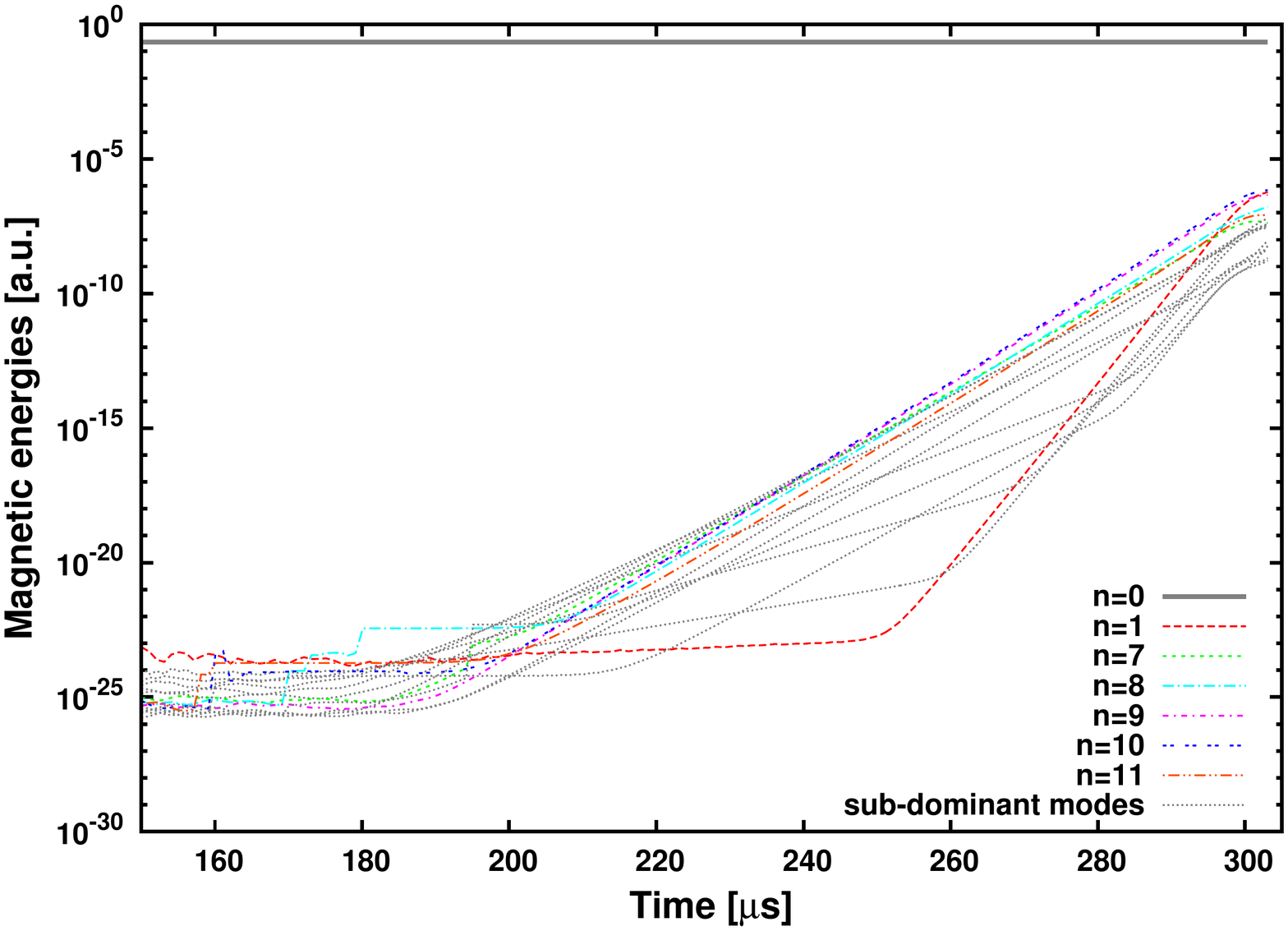}
\caption{Time-traces of the magnetic energies contained in the individual toroidal modes
are shown for the simulation with periodicity $1$. For clarity, kinetic energies are omitted and
sub-dominant modes are only indicated by dotted gray lines.
It is remarkable that the $n=1$ mode
reaches a comparable energy level at the onset of non-linear saturation
as the $n=10$ mode, which is the linearly most unstable mode.}
\label{fig:energies_ntor33}
\end{figure}

Time-traces of the magnetic energies contained in each toroidal harmonic
are shown in Figure~\ref{fig:energies_ntor33}. Linearly, the $n=10$ mode
has the largest growth rate $\gamma_\text{SI}\approx\unit{2.0\times10^5}{\second^{-1}}$.
In a similar way as described for the $n=16$ mode in the previous Section,
the initially very small growth rate of the $n=1$ mode
($\gamma_\text{SI}\approx\unit{2\times10^4}{\second^{-1}}$) suddenly
changes at $t=\unit{150}{\micro\second}$ due to the non-linear
interaction between the toroidal harmonics and becomes very large:
$\gamma_\text{SI}\approx\unit{4\times10^5}{\second^{-1}}$. In the non-linear
phase of the mode, the $n=1$ perturbation reaches a similar magnetic energy
as the $n=10$ perturbation which remains dominant also at the beginning of non-linear
saturation ($t\approx\unit{300}{\micro\second}$).
A first important effect that cannot be covered in simulations with
low toroidal resolution (i.e., high periodicity) is that low-$n$ modes can grow
to large amplitudes non-linearly. The growth rate of the dominant mode ($n=10$ in
our case) is not affected significantly by the toroidal mode-coupling.
Also, the radial propagation velocity of the filaments into the vacuum
region hardly changes compared to the case with low toroidal resolution:
The filaments accelerate in the beginning and saturate at a velocity
of about $\unit{3}{\kilo\meter\per\second}$.

\begin{figure}
\centering
  \includegraphics[height=0.5\textwidth]{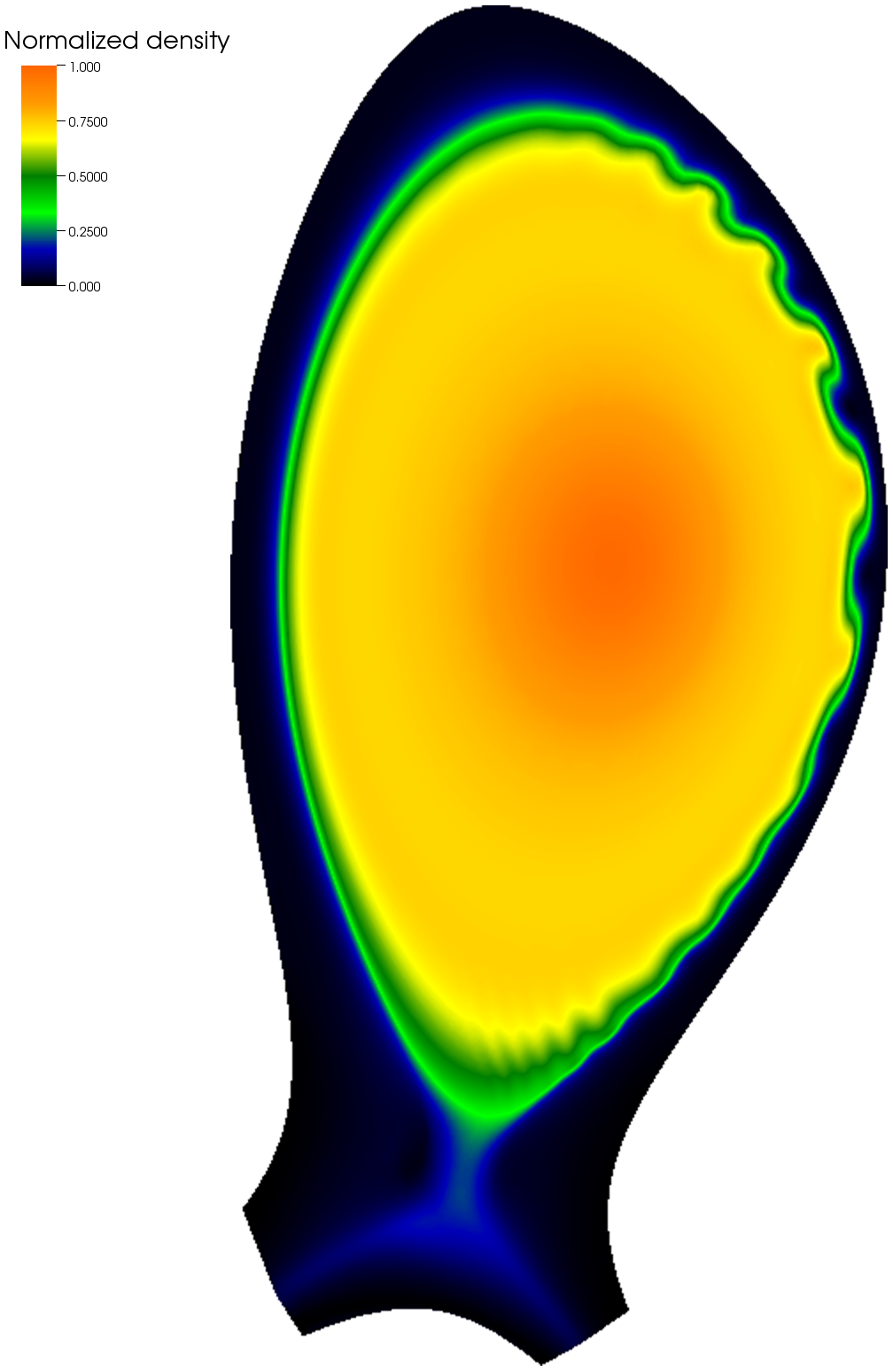}
\caption{The density-perturbation observed in the simulation with periodicity $1$
is plotted at $\unit{294}{\micro\second}$. The ballooning-structures become less regular and perturbations are
strong only within a localized region. In the cross-section shown, this region
is located on the upper low-field side.}
\label{fig:density2}
\end{figure}

The developing density perturbation is shown
in Figure~\ref{fig:density2}. Also with high toroidal resolution, a
ballooning-like structure is produced at the low-field side of the plasma.
The poloidal size of the ballooning-fingers is around $\unit{10-12}{\centi\meter}$
at the midplane. In comparison to simulations with low toroidal resolution,
these structures are a bit smaller. A significant difference becomes obvious
when comparing Figures~\ref{fig:density1} and~\ref{fig:density2}:
Due to the mode-coupling, not all fingers grow to the same amplitude. A cluster
of fingers can be seen that develops much stronger than the rest of the
ballooning-structures.
A strong localization of perturbations has also been
observed in a ballooning-instability simulated with the BOUT code~\citep{Snyder2005}.


\begin{figure}
\centering
  \includegraphics[width=0.9\textwidth]{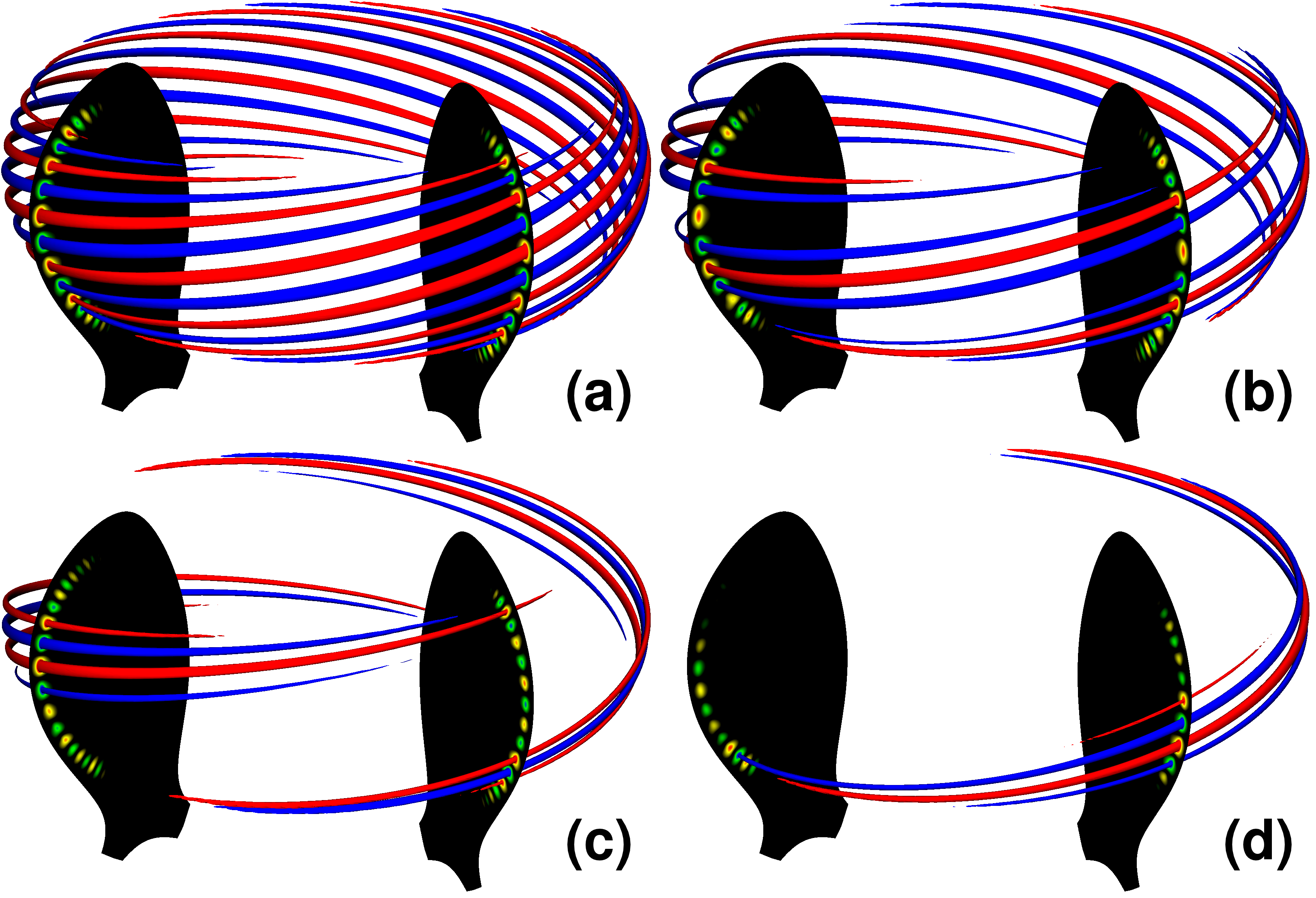}
\caption{Contours of the poloidal flux perturbation are shown for simulations
with \textbf{(a)} periodicity $8$, \textbf{(b)} $4$, \textbf{(c)} $2$,
and \textbf{(d)} $1$, respectively. The red and blue contours
are plotted at the surfaces corresponding to the perturbed poloidal flux values
$\tilde{\Psi}_\text{red/blue}=\pm0.7\cdot(|\tilde{\Psi}_\text{min}|+\tilde{\Psi}_\text{max})/2$. Here,
$\tilde{\Psi}_\text{min}$ and $\tilde{\Psi}_\text{max}$ denote the strongest negative 
positive perturbation values, respectively. At lower periodicities, the perturbation
steadily gets more localized.}
\label{fig:flux-localisation}
\end{figure}

\begin{figure}
\centering
  \includegraphics[width=0.45\textwidth]{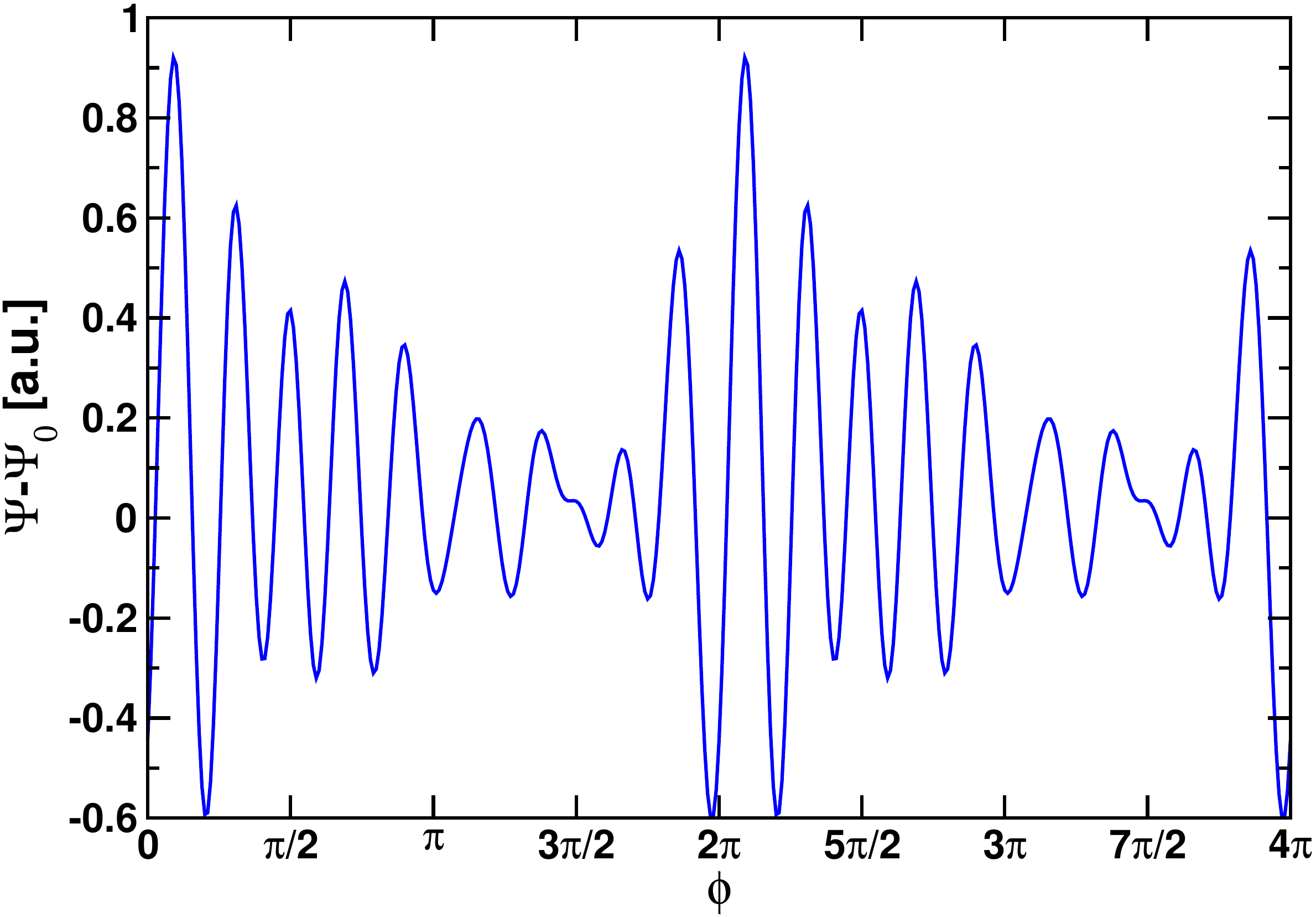}
\caption{The perturbation of the poloidal flux at the outboard midplane is shown
for the simulation with periodicity $1$ 
versus the toroidal angle for two transits around the torus. The perturbation
amplitude shows a strong toroidal variation equivalent to a localization of the
perturbation to $\Delta\phi\approx\unit{3}{\rad}$ (f.w.h.m.). As equilibrium,
boundary conditions, and sources are completely axi-symmetric, the localization
position is essentially arbitrary which proves to be true when looking at a set
of different simulations.}
\label{fig:psi-vs-phi}
\end{figure}

The localization of the perturbation becomes even more obvious
when the magnetic footprint of the mode is considered. in
Figure~\ref{fig:flux-localisation}, the perturbation of the poloidal
magnetic flux is plotted for simulations with different periodicities.
Clearly, the localization of the mode
can only be described correctly in simulations with periodicity $1$.
Figure~\ref{fig:psi-vs-phi} shows the perturbation of the poloidal flux at the
outboard midplane versus the toroidal angle.

\begin{figure}
\centering
  \includegraphics[height=0.5\textwidth]{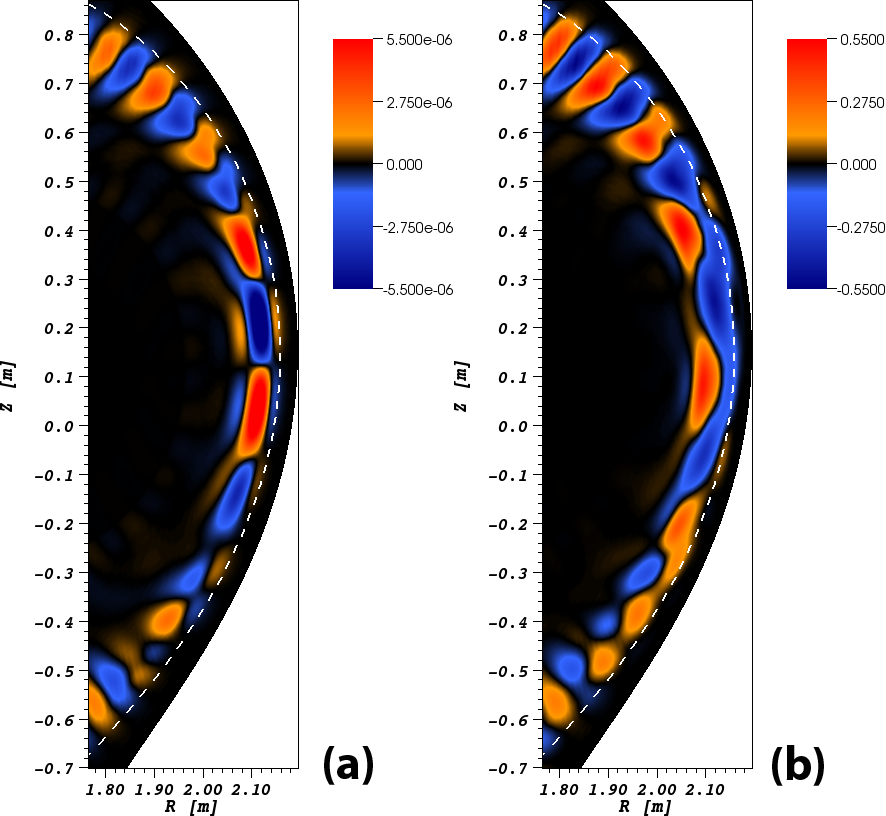}
\caption{The current perturbation at the plasma edge is shown for the
simulation with periodicity $1$ in the \textbf{(a)} linear
($\unit{240}{\micro\second}$) and \textbf{(b)} non-linear phases
($\unit{300}{\micro\second}$). In the non-linear phase, large
regions with positive respectively negative current (this cross-section)
form at the separatrix (dashed line) around the midplane.}
\label{fig:current-perturbation}
\end{figure}

The perturbation is already localized in the linear phase of the mode. A
qualitative change between the linear and the non-linear phases is shown
in Figure~\ref{fig:current-perturbation}, where the current perturbation is
plotted for two different time-frames in the simulation with periodicity $1$.
In the non-linear phase where the ballooning-fingers become visible
in the density perturbation, the previously alternating current filaments
merge at the position of the separatrix around the outer midplane. Large areas
with positive respectively negative currents are created.


\begin{figure}
\centering
  \includegraphics[width=0.6\textwidth]{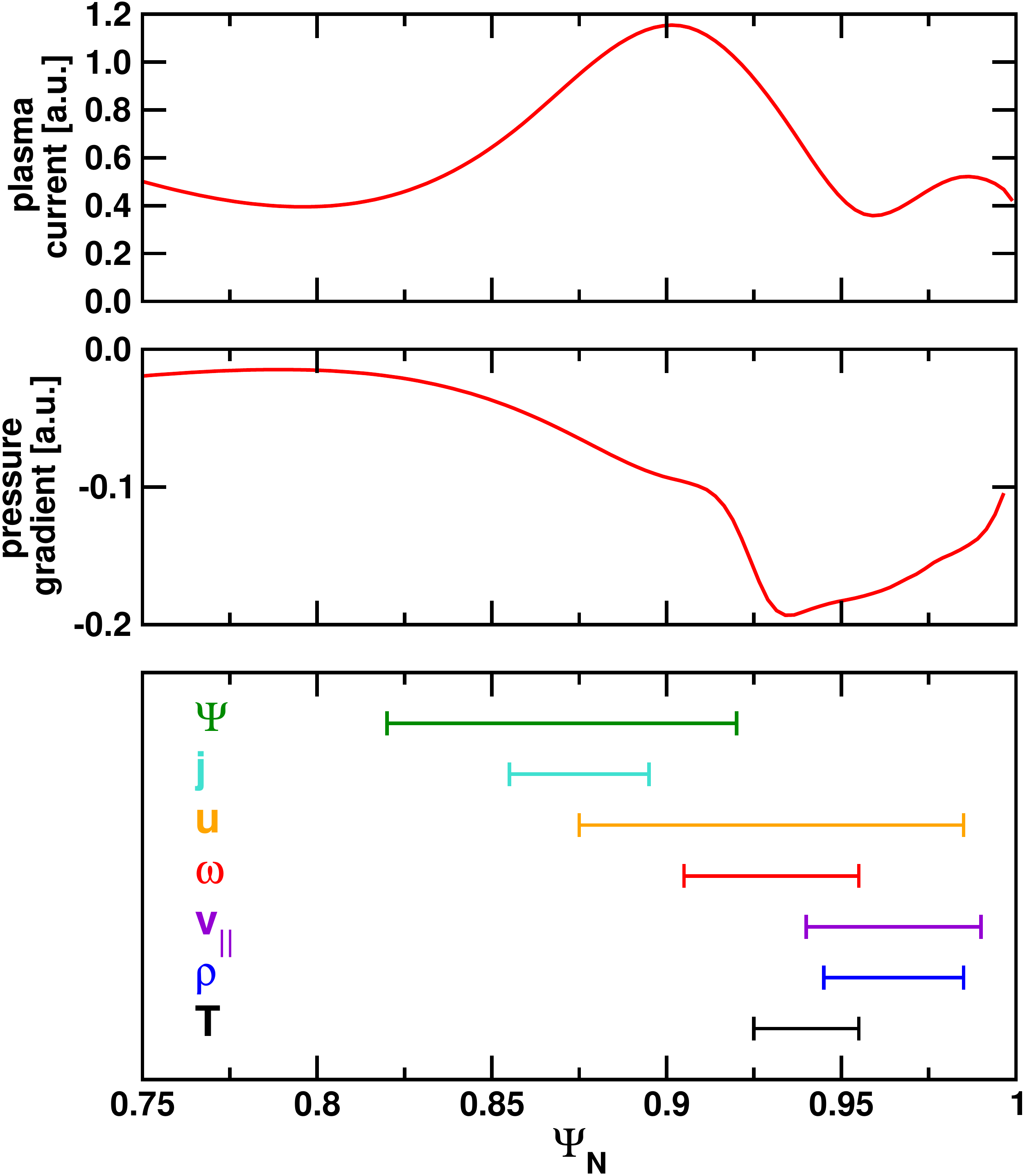}
\caption{For the simulation with periodicity $1$, the radial positions of the strongest
perturbations are shown at $t=\unit{300}{\micro\second}$ for the seven physical variables
and are compared to profiles of the plasma current and the pressure gradient.}
\label{fig:radial-location}
\end{figure}

The strongest perturbations of all physical quantities are localized in a
flux-tube like region which extends from the vicinity of the lower
active X-point along magnetic field lines to the vicinity of the upper
inactive X-point (compare Figure~\ref{fig:flux-localisation}). The
perturbations are strongest around the midplane. As an exception,
$v_{||}$ is perturbed especially around the end-points of this flux-tube,
a consequence of field-line stagnation close to the X-points.
However, the radial perturbation positions differ as shown in
Figure~\ref{fig:radial-location}. It can be seen, that the strongest
perturbations of the magnetic quantities $\Psi$ and $j$ are located in the
region of strong plasma current, while the kinetic quantities are
perturbed further outwards in the region of strong pressure
gradients.

\subsection{Towards more Realistic Resistivities}\label{:results:loweta}

This Section briefly shows results for simulations with the plasma
viscosity and resistivity reduced by a factor of $10$ (\texttt{eta6} cases)
compared to the simulations shown above. To be able to resolve
these more realistic parameters, the number of 2D Bezier elements in
the poloidal plane was increased by a factor of two. These simulations
need to be considered with care as the most unstable mode
is $n=13$ while we do not take into account mode numbers beyond $n=16$
for computational reasons.

\begin{figure}
\centering
  \includegraphics[width=0.9\textwidth]{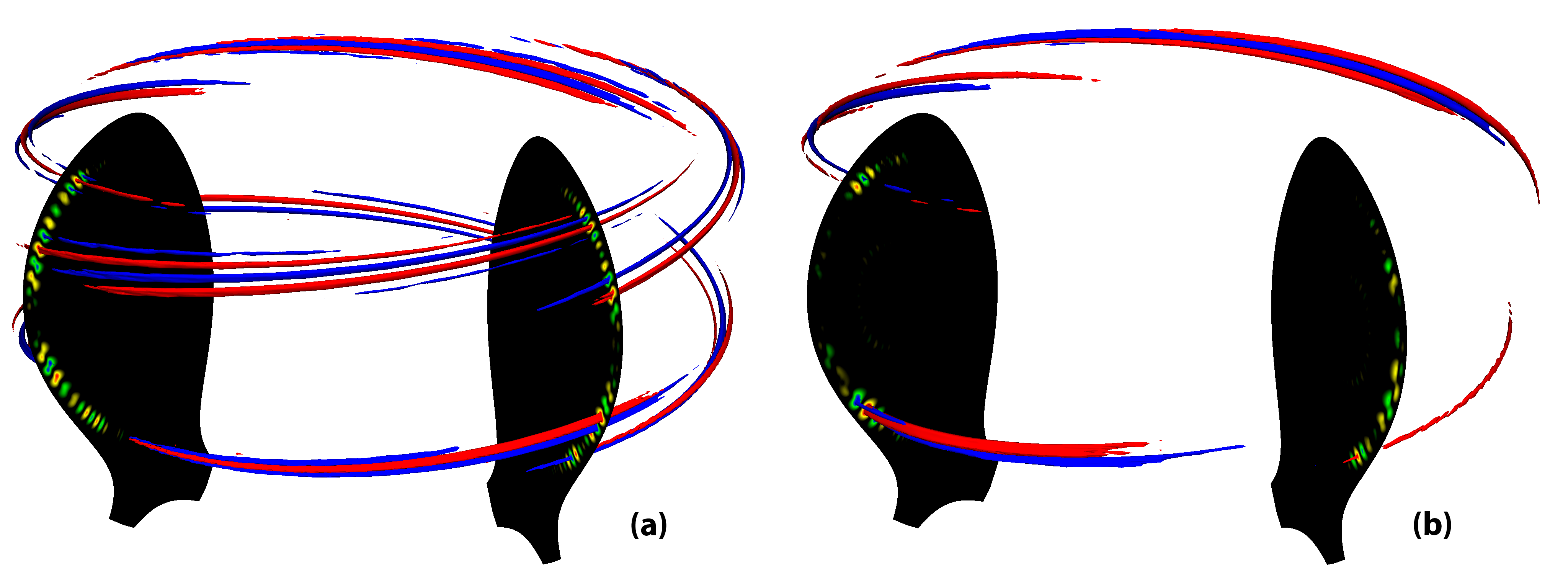}
\caption{For simulations with \textbf{(a)} periodicity $2$
respectively \textbf{(b)} $1$ where the
plasma resistivity and viscosity is reduced by a factor of $10$
compared to the simulations presented above, the poloidal flux
perturbation is shown analogously to Figure~\ref{fig:flux-localisation}.
A strong localization of perturbations is observed in these simulations as well.}
\label{fig:flux-localisation2}
\end{figure}

It can be seen that a strong localization of the perturbations is observed
at periodicity $1$ as in the \texttt{eta5}-cases. This is shown for the
perturbation of the poloidal flux in Figure~\ref{fig:flux-localisation2}.
In contrast to the \texttt{eta5}-simulations, the perturbation maximum is
not located around the midplane but more towards the top and bottom regions
of the low-field side. This distribution of the flux-perturbation is not
an artifact caused by cutting toroidally at $n=16$: A simulation with periodicity
$2$ was carried out in which the toroidal modes $n=0,2,\dots,20,22$ are
resolved, where the strongest perturbation of the poloidal flux
is not observed at the midplane but above and below it, as well.

At ASDEX Upgrade, an off-midplane mode-structure has recently been observed
in the temperature using ECE-Imaging~\citep{Boom2011}.
In our simulations, the perturbation maximum of the kinetic quantities is,
however, located around the midplane. This is a consequence of the comparably
large plasma resistivities in our simulations which allow magnetic and kinetic
quantities to decouple. At smaller resistivities, which we cannot resolve at
present, also the kinetic quantities might show an off-midplane behavior.

\section{Comparison to Experiments}\label{:comparison}

\begin{figure}
\centering
  \includegraphics[width=0.9\textwidth]{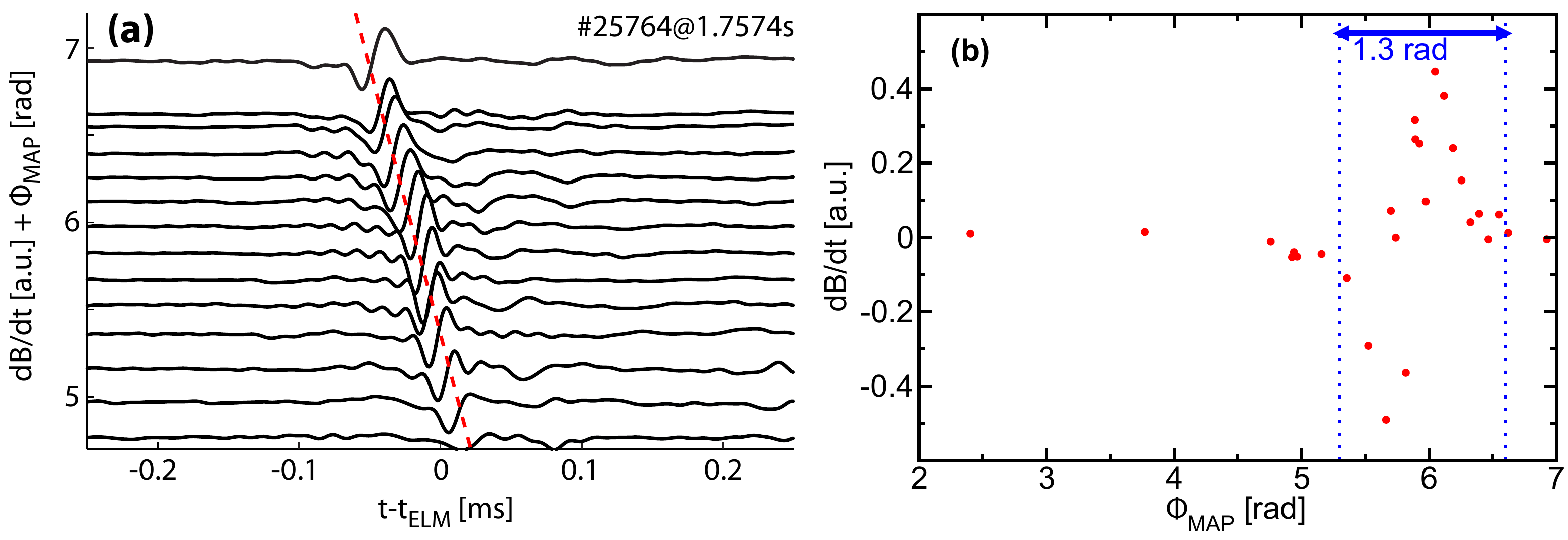}
\caption{\textbf{(a)} Magnetic signals from selected
Mirnov-coils are shown for a strongly localized 
solitary magnetic perturbation in ASDEX Upgrade as in Reference~\citep{WENNINGER2011}.
All measurement locations are mapped to a common toroidal angle $\phi_\text{MAP}$ via
field-line tracing. The magnetic perturbation propagates with constant toroidal velocity
in the electron diamagnetic drift direction in the lab-frame as indicated by the red
dashed line. The onset time of the erosion of pedestal temperature and density profiles
is denoted $t_\text{ELM}$. \textbf{(b)} The time-derivative of the magnetic field measured
by Mirnov coils is plotted versus the toroidal mapping angle $\phi_\text{MAP}$ at
$t-t_\text{ELM}=\unit{-0.03}{\milli\second}$. The solitary magnetic perturbation is
localized to $\Delta\phi\approx\unit{1.2}{\rad}$.}
\label{fig:smp-exp}
\end{figure}

In this Section, some properties of the simulated edge instabilities
are compared to experimental observations. This shows that important
aspects of the early phase of edge localized modes are well described by
the reduced MHD model. More detailed comparisons between JOREK simulations
of complete ELM crashes and experimental measurements at ASDEX Upgrade
are planned for the future (e.g., evolution of pedestal gradients, detachment
of filaments, heat-flux patterns at divertor plates).

The poloidal flux perturbation from the simulation with periodicity $1$
shown in Figure~\ref{fig:psi-vs-phi} exhibits a toroidal localization: 
Large perturbation amplitudes are localized to a region of about
$\Delta\phi\approx\unit{3}{\rad}$.
Thus, the modes we observe in our simulations of the early ELM phase when
simulating the full torus (periodicity $1$) exhibit a similar magnetic
structure as so-called solitary magnetic perturbations recently discovered at
the ELM onset in ASDEX Upgrade and described in great detail in
Reference~\citep{WENNINGER2011}. From the systematic analysis of a large
number of ELM crashes, a continuous distribution of the mode solitariness was
reported between cases with a very pronounced toroidal localization
(an example is shown in Figure~\ref{fig:smp-exp}) and cases with a magnetic
perturbation strength that is toroidally virtually uniform. The toroidal localization
observed in our simulations (localized to $\Delta\phi\approx\unit{3}{\rad}$) is
less pronounced than the extreme example of Figure~\ref{fig:smp-exp}b with
$\Delta\phi\approx\unit{1.3}{\rad}$. A direct comparison is planned for the
future making use of a virtual magnetic diagnostic which determines
magnetic signals from the simulations at the same positions as the Mirnov
coils.
Toroidally asymmetric structures at ELMs are also described from experimental
observations in
References~\citep{Reimerdes1998,Becoulet2003,Eich2003,Bobkov2004}. In analytical calculations,
localized instabilities were also reported by Wilson et.al.~\citep{Wilson2004}.
These ``explosive ballooning'' instabilities grow much faster non-linearly
than linearly and a poloidal narrowing of
the instability in the non-linear phase is reported. Both features are
not observed in the simulated edge instabilities which indicates that different
mechanisms are responsible for the localization in our simulations.

The dominant toroidal mode-number turns out to be $10$ in the simulations.
With the plasma resistivity reduced towards more realistic values, the
dominant mode number shifts towards $13$. 
This is in quite good agreement to experimental findings in the tokamaks
ASDEX Upgrade and MAST for type-I ELMs, where mode-numbers of $8-24$ were
observed in energy deposition patterns~\citep{Eich2003}, around $15$ was
found from measurements with the midplane manipulator
and visible-light imaging~\citep{Kirk2005}, and mode numbers of $18\pm4$ have
been obtained for the onset of the ELM-crash using
ECE-Imaging~\citep{Boom2011}. Uncertainties in our simulations
come from the limited poloidal resolution and the neglect of diamagnetic
stabilization as discussed in Section \ref{:simulations}.

Low-$n$ modes gain large amounts of energy non-linearly in our simulations
with periodicity $1$. This allows them to interact much more efficiently
with core-MHD modes like tearing modes which typically also feature low
toroidal mode numbers like $1$ or $2$. Indeed, there is experimental evidence
from the DIII-D tokamak that ELMs are an important triggering mechanism
for neoclassical tearing modes~\citep{Sauter1997}.

The poloidal extent of filaments on the outer midplane observed in
simulations with high toroidal resolution is around $\unit{10-12}{\centi\meter}$.
Measurements in ASDEX Upgrade and MAST revealed filament sizes perpendicular
to the field lines of $\unit{5-10}{\centi\meter}$~\citep{Kirk2005}. For
ASDEX Upgrade, perpendicular and poloidal filament sizes are equivalent
due to the small field-line pitch-angle such that simulation results and
experimental observations show good agreement here as well.

In the simulations, the radial velocity of the developing finger structures
saturates at about \unit{3}{\kilo\meter\per\second} after an initial
acceleration. This corresponds to a distribution of the radial filament speed
with an upper limit of \unit{3}{\kilo\meter\per\second}.
This velocity depends on the stability of the initial
equilibrium. The unrealistically large values for the plasma resistivity
might lead to an over-estimation of the filament speeds, while the ideal-wall
boundary conditions tend to reduce the radial velocity.
In experimental measurements, the radial filament speed is
found to be distributed around \unit{1}{\kilo\meter\per\second} in ASDEX
Upgrade~\citep{Schmid2008,Kirk2011}. Filament speeds faster
than \unit{2}{\kilo\meter\per\second} occur in 20\% of the cases in both
References and almost no filaments faster than \unit{3}{\kilo\meter\per\second}
are observed. Hence, radial filament speeds in simulations and
experimental measurements seem to agree reasonably well.

In the magnetic quantities, an off-midplane mode-structure is observed in
the simulations with lower plasma resistivity (\texttt{eta6} simulations). As
the resistivity is still unrealistically large in these simulations, magnetic
and kinetic quantities are decoupled such that the strongest perturbation of
the temperature is located at the midplane. Still, this might be related to
the off-midplane structures observed by ECE-Imaging in ASDEX
Upgrade~\citep{Boom2011}.

\section{Conclusions and Outlook}\label{:conclusions}

Exponentially growing ballooning-like modes have been simulated with the
reduced-MHD version of the non-linear MHD code JOREK
in the geometry and using the profiles of a typical ASDEX Upgrade
H-mode discharge. Dominant toroidal mode numbers, poloidal filament sizes,
and radial filament-propagation speeds of these instabilities are in
good agreement with experimental observations for type-I ELMs in ASDEX Upgrade.
At sufficient toroidal resolution, perturbations show a pronounced toroidal
and poloidal localization which is compatible with solitary magnetic perturbations
recently discovered in ASDEX Upgrade. In some cases, the perturbation
of the magnetic flux is stronger at the top and bottom low-field side than at
the midplane. Presumably due to a decoupling of magnetic and kinetic quantities caused by
the unrealistically large plasma resistivity, density and temperature
perturbations are always localized on the midplane of the low-field side.
Strong perturbations in the low-$n$ modes are triggered non-linearly
in the simulations with periodicity $1$ and might explain the strong interaction
of ELMs with core-MHD modes like neoclassical tearing modes observed in some
experiments.

While this work concentrates on the early phase of an ELM, further studies are
planned to compare the simulation of a full ELM crash to experimental observations
requiring a more sophisticated modeling of the scrape-off layer. Simulations of
a full ELM cycle will also be attempted. Future numerical improvements and
increased computational resources will be used to advance our investigations
towards more realistic plasma parameters while keeping high toroidal resolutions.

\section{Acknowledgments}\label{:ack}

The simulations were mostly carried out on the HPC-FF computing
cluster in J\"ulich, Germany. One of the authors~(K.L.) would like
to acknowledge the support by the Austrian Science Fund~(FWF) under
grant No.\ P19901. Additionally, the authors would like to thank
Erika Strumberger and Mike Dunne for their help with the plasma
equilibrium reconstruction and Klaus Reuter for assistance with
visualizations.

\appendix

\section{Derivation of the Induction Equation}\label{:app:inductioneq}

The reduced MHD equations implemented in the JOREK code can be derived
following Reference~\citep{Strauss1997}. For the induction equation,
this is shown in the following. The starting points are the well-known
expression for the electric field in the MHD approximation,
\begin{equation}
  \mathbf{E}=-\mathbf{v}\times\mathbf{B} + \eta\mathbf{j}  \label{eq:app10}
\end{equation}
and the Maxwell-Faraday law expressed in the vector potential $\mathbf{A}$,
\begin{equation}
  \pderiv{\mathbf{A}}{t} = -\mathbf{E} - F_0\nabla u.  \label{eq:app20}
\end{equation}
Inserting \eqref{eq:app10} into \eqref{eq:app20} and
multiplying it with the toroidal unity vector $\hat{\mathbf{e}}_\phi$ yields
\begin{equation}
  \pderiv{\Psi}{t} = \eta j + (\mathbf{v}\times\mathbf{B})\cdot\hat{\mathbf{e}}_\phi
    - F_0\nabla u\cdot\hat{\mathbf{e}}_\phi,\label{eq:app30}
\end{equation}
where the poloidal flux is identified as the major radius times the toroidal component
of the vector potential,
$\Psi\equiv R\;\mathbf{A}\cdot\hat{\mathbf{e}}_\phi$, and
$j=-\mathbf{j}\cdot\hat{\mathbf{e}}_\phi$ denotes the toroidal plasma current. Using
Equations~\eqref{eq:v-def} and~\eqref{eq:B-def}, this can be written as
\begin{equation}
  \pderiv{\Psi}{t} = \eta j - R[u,\Psi] - F_0\pderiv{u}{\phi},\label{eq:app40}
\end{equation}
which is the induction equation (Equation~\eqref{eq:psievol}) solved in the JOREK
reduced MHD model with the Poisson bracket
$[u,\Psi]=\pderiv{u}{R}\pderiv{\Psi}{Z}-\pderiv{u}{Z}\pderiv{\Psi}{R}$.
In the last step, the reduced MHD approximation to first order in 
$\varepsilon=\nabla_{||}\Psi/\nabla_\bot\Psi\ll1$ yielding
$\mathbf{v}_\text{pol}\approx\mathbf{v}_\bot$ was applied. 

The poloidal components of Equation~\eqref{eq:app20}, obtained by applying the
operator $\hat{\mathbf{e}}_\phi\times$ to this equation, yield a definition equation
for the poloidal velocity (see poloidal components of Equation~\eqref{eq:v-def}) in
which $u$ can be identified as the poloidal velocity stream function. In this set of
equations, $u$ also acts as electric potential (except for a constant factor $F_0$).

Galilei-invariance of the induction equation (Equation~\eqref{eq:app20}) is not
obvious at first glance. However, the proof is straightforward when taking into
account that the scalar potential $\phi=F_0 u$ is modified according to
$\phi\rightarrow\phi-\mathbf{v}_0\cdot\mathbf{A}$ under the transformation
$\mathbf{v}\rightarrow\mathbf{v}-\mathbf{v}_0$ while the vector potential
remains unchanged (non-relativistic limit). In the large aspect-ratio limit,
it can also be shown easily that the reduced-MHD induction equation
(Equation~\eqref{eq:app40}) is invariant
to a transformation $\mathbf{v}\rightarrow\mathbf{v}-\mathbf{v}_z$ with $z$
along the cylinder axis, as the scalar potential transforms
according to $\phi\rightarrow\phi-v_z\Psi$.


\begin{thebibliography}{33}
\providecommand{\natexlab}[1]{#1}
\providecommand{\url}[1]{\texttt{#1}}
\expandafter\ifx\csname urlstyle\endcsname\relax
  \providecommand{\doi}[1]{doi: #1}\else
  \providecommand{\doi}{doi: \begingroup \urlstyle{rm}\Url}\fi

\bibitem[Loarte et~al.(2003)Loarte, Saibene, Sartori, Campbell, Becoulet,
  Horton, Eich, Herrmann, Matthews, Asakura, Chankin, Leonard, Porter,
  Federici, Janeschitz, Shimada, and Sugihara]{Loarte2003}
A~Loarte, G~Saibene, R~Sartori, D~Campbell, M~Becoulet, L~Horton, T~Eich,
  A~Herrmann, G~Matthews, N~Asakura, A~Chankin, A~Leonard, G~Porter,
  G~Federici, G~Janeschitz, M~Shimada, and M~Sugihara.
\newblock Characteristics of type {I} {ELM} energy and particle losses in
  existing devices and their extrapolation to {ITER}.
\newblock \emph{Plasma Physics and Controlled Fusion}, 45\penalty0
  (9):\penalty0 1549, 2003.
\newblock URL \url{http://stacks.iop.org/0741-3335/45/i=9/a=302}.

\bibitem[Klimov et~al.(2011)Klimov, Podkovyrov, Zhitlukhin, Kovalenko, Linke,
  Pintsuk, Landman, Pestchanyi, Bazylev, Janeschitz, Loarte, Merola, Hirai,
  Federici, Riccardi, Mazul, Giniyatulin, Khimchenko, and Koidan]{Klimov2011}
N.~Klimov, V.~Podkovyrov, A.~Zhitlukhin, D.~Kovalenko, J.~Linke, G.~Pintsuk,
  I.~Landman, S.~Pestchanyi, B.~Bazylev, G.~Janeschitz, A.~Loarte, M.~Merola,
  T.~Hirai, G.~Federici, B.~Riccardi, I.~Mazul, R.~Giniyatulin, L.~Khimchenko,
  and V.~Koidan.
\newblock Experimental study of {PFCs} erosion and eroded material deposition
  under {ITER}-like transient loads at the plasma gun facility {QSPA-T}.
\newblock \emph{Journal of Nuclear Materials}, 415\penalty0 (1,
  Supplement):\penalty0 S59 -- S64, 2011.
\newblock ISSN 0022-3115.
\newblock \doi{10.1016/j.jnucmat.2011.01.013}.
\newblock URL
  \url{http://www.sciencedirect.com/science/article/pii/S0022311511000250}.
\newblock <ce:title>Proceedings of the 19th International Conference on
  Plasma-Surface Interactions in Controlled Fusion</ce:title>.

\bibitem[Huysmans and Czarny(2007)]{Huysmans2007}
G.T.A. Huysmans and O.~Czarny.
\newblock {MHD} stability in {X}-point geometry: simulation of {ELMs}.
\newblock \emph{Nuclear Fusion}, 47\penalty0 (7):\penalty0 659, 2007.
\newblock URL \url{http://stacks.iop.org/0029-5515/47/i=7/a=016}.

\bibitem[Huysmans et~al.(2008)Huysmans, Abgrall, Becoulet, Huart, Nkonga,
  Pamela, van~der Plas, and Ramet]{Huysmans2008}
G.T.A. Huysmans, R.~Abgrall, M.~Becoulet, R.~Huart, B.~Nkonga, S.~Pamela,
  E.~van~der Plas, and P.~Ramet.
\newblock Non-linear {MHD} simulations of {ELMs}.
\newblock In \emph{Proceedings of the 35th EPS Conference on Plasma Physics},
  pages P--2.065, Hersonissos, 2008.
\newblock URL \url{epsppd.epfl.ch/Hersonissos/pdf/P2_065.pdf}.

\bibitem[Huysmans et~al.(2009)Huysmans, Pamela, van~der Plas, and
  Ramet]{Huysmans2009}
G~T~A Huysmans, S~Pamela, E~van~der Plas, and P~Ramet.
\newblock Non-linear {MHD} simulations of edge localized modes ({ELMs}).
\newblock \emph{Plasma Physics and Controlled Fusion}, 51\penalty0
  (12):\penalty0 124012, 2009.
\newblock URL \url{http://stacks.iop.org/0741-3335/51/i=12/a=124012}.

\bibitem[Pamela et~al.(2010)Pamela, Huysmans, and Benkadda]{Pamela2010}
S~Pamela, G~Huysmans, and S~Benkadda.
\newblock Influence of poloidal equilibrium rotation in {MHD} simulations of
  edge-localized modes.
\newblock \emph{Plasma Physics and Controlled Fusion}, 52\penalty0
  (7):\penalty0 075006, 2010.
\newblock URL \url{http://stacks.iop.org/0741-3335/52/i=7/a=075006}.

\bibitem[Huysmans et~al.(2010)Huysmans, Pamela, Beurskens, Becoulet, and
  van~der Plas]{Huysmans2010}
G.~Huysmans, S.~Pamela, M.~Beurskens, M.~Becoulet, and E.~van~der Plas.
\newblock Non-linear {MHD} simulation of natural and pellet triggered {ELMs}.
\newblock In \emph{Proceedings of th 23rd IAEA Fusion Energy Conference},
  Daejon, South Korea, 2010.

\bibitem[Pamela et~al.(2011)Pamela, Huysmans, Beurskens, Devaux, Eich,
  Benkadda, and contributors]{Pamela2011}
S~J~P Pamela, G~T~A Huysmans, M~N~A Beurskens, S~Devaux, T~Eich, S~Benkadda,
  and {JET}~{EFDA} contributors.
\newblock Nonlinear {MHD} simulations of edge-localized-modes in {JET}.
\newblock \emph{Plasma Physics and Controlled Fusion}, 53\penalty0
  (5):\penalty0 054014, 2011.
\newblock URL \url{http://stacks.iop.org/0741-3335/53/i=5/a=054014}.

\bibitem[H\"olzl et~al.(2011)H\"olzl, G\"unter, and {ASDEX Upgrade
  Team}]{Hoelzl2011}
M.~H\"olzl, S.~G\"unter, and {ASDEX Upgrade Team}.
\newblock Reduced-{MHD} simulations of edge localized modes in {ASDEX Upgrade}
  ({P2.078}).
\newblock In \emph{Proceedings of the 38th EPS Conference on Plasma Physics},
  Strasbourg, France, 2011.

\bibitem[Dudson et~al.(2011)Dudson, Xu, Umansky, Wilson, and
  Snyder]{Dudson2011}
B~D Dudson, X~Q Xu, M~V Umansky, H~R Wilson, and P~B Snyder.
\newblock Simulation of edge localized modes using {BOUT++}.
\newblock \emph{Plasma Physics and Controlled Fusion}, 53\penalty0
  (5):\penalty0 054005, 2011.
\newblock URL \url{http://stacks.iop.org/0741-3335/53/i=5/a=054005}.

\bibitem[Pankin et~al.(2007)Pankin, Bateman, Brennan, Kritz, Kruger, Snyder,
  Sovinec, and the {NIMROD}~team]{Pankin2007}
A~Y Pankin, G~Bateman, D~P Brennan, A~H Kritz, S~Kruger, P~B Snyder, C~Sovinec,
  and the {NIMROD}~team.
\newblock Modelling of {ELM} dynamics for {DIII-D} and {ITER}.
\newblock \emph{Plasma Physics and Controlled Fusion}, 49\penalty0
  (7):\penalty0 S63, 2007.
\newblock URL \url{http://stacks.iop.org/0741-3335/49/i=7/a=S04}.

\bibitem[Sugiyama and Strauss(2010)]{Sugiyama2010}
L.~E. Sugiyama and H.~R. Strauss.
\newblock Magnetic {X}-points, edge localized modes, and stochasticity.
\newblock \emph{Physics of Plasmas}, 17\penalty0 (6):\penalty0 062505, 2010.
\newblock \doi{10.1063/1.3449301}.
\newblock URL \url{http://link.aip.org/link/?PHP/17/062505/1}.

\bibitem[Czarny and Huysmans(2008)]{Czarny2008}
O~Czarny and G~Huysmans.
\newblock Bezier surfaces and finite elements for {MHD} simulations.
\newblock \emph{Journal of Computational Physics}, 227\penalty0 (16):\penalty0
  7423 -- 7445, 2008.
\newblock ISSN 0021-9991.
\newblock \doi{10.1016/j.jcp.2008.04.001}.
\newblock URL
  \url{http://www.sciencedirect.com/science/article/pii/S0021999108002118}.

\bibitem[Hermann(2003)]{HERMANN03A}
A.~Hermann.
\newblock Special issue: {ASDEX Upgrade}.
\newblock \emph{Fusion Science and Technology}, 44\penalty0 (3), 2003.
\newblock URL \url{http://www.new.ans.org/pubs/journals/fst/v_44:3}.

\bibitem[Wolfrum et~al.(2011)Wolfrum, Bernert, Boom, Burckhart, Classen,
  Conway, Eich, Fischer, Gude, Herrmann, Jr, Maraschek, McDermott, Park,
  Puetterich, Vicente, Wieland, Willensdorfer, and the {ASDEX Upgrade
  Team}]{Wolfrum2011}
E~Wolfrum, M~Bernert, J~E Boom, A~Burckhart, I~G~J Classen, G~D Conway, T~Eich,
  R~Fischer, A~Gude, A~Herrmann, N~C~Luhmann Jr, M~Maraschek, R~McDermott, H~K
  Park, T~Puetterich, J~Vicente, B~Wieland, M~Willensdorfer, and the {ASDEX
  Upgrade Team}.
\newblock Characterization of edge profiles and fluctuations in discharges with
  type-{II} and nitrogen-mitigated edge localized modes in {ASDEX Upgrade}.
\newblock \emph{Plasma Physics and Controlled Fusion}, 53\penalty0
  (8):\penalty0 085026, 2011.
\newblock URL \url{http://stacks.iop.org/0741-3335/53/i=8/a=085026}.

\bibitem[Strauss(1997)]{Strauss1997}
H.~R. Strauss.
\newblock Reduced mhd in nearly potential magnetic fields.
\newblock \emph{Journal of Plasma Physics}, 57\penalty0 (1):\penalty0 83--87,
  1997.
\newblock \doi{10.1017/S0022377896005296}.

\bibitem[Stangeby and Chankin(1995)]{Stangeby1995}
P.~C. Stangeby and A.~V. Chankin.
\newblock The ion velocity ({Bohm}--{Chodura}) boundary condition at the
  entrance to the magnetic presheath in the presence of diamagnetic and {E} x
  {B} drifts in the scrape-off layer.
\newblock \emph{Physics of Plasmas}, 2\penalty0 (3):\penalty0 707--715, 1995.
\newblock \doi{10.1063/1.871421}.
\newblock URL \url{http://link.aip.org/link/?PHP/2/707/1}.

\bibitem[Hirsch(1989)]{Hirsch1989}
Charles Hirsch.
\newblock \emph{Numerical Computation of Internal and External Flows, Volume 1,
  Fundamentals of Numerical Discretization}.
\newblock Wiley, 1989.
\newblock ISBN 978-0-471-92385-5.

\bibitem[PAS()]{PASTIX}
Website of {PASTIX}.
\newblock URL \url{https://gforge.inria.fr/projects/pastix/}.

\bibitem[McCarthy et~al.(1999)McCarthy, Martin, and Schneider]{McCarthy1999}
P.~J. McCarthy, P.~Martin, and W.~Schneider.
\newblock The {CLISTE} interpretive equilibrium code.
\newblock Technical Report IPP-Report 5/85, Max-Planck-Institut f\"ur
  Plasmaphysik, Boltzmannstra\ss e 2, 85748 Garching, Germany, 1999.
\newblock URL
  \url{http://edoc.mpg.de/display.epl?mode=doc&id=413804&col=33&grp=1311#cb}.

\bibitem[McCarthy(1999)]{McCarthy1999a}
P.~J. McCarthy.
\newblock Analytical solutions to the {Grad-Shafranov} equation for tokamak
  equilibrium with dissimilar source functions.
\newblock \emph{Phys. Plasmas}, 6\penalty0 (9):\penalty0 3554, 1999.
\newblock \doi{10.1063/1.873630}.

\bibitem[Snyder et~al.(2005)Snyder, Wilson, and Xu]{Snyder2005}
P.~B. Snyder, H.~R. Wilson, and X.~Q. Xu.
\newblock Progress in the peeling-ballooning model of edge localized modes:
  Numerical studies of nonlinear dynamics.
\newblock \emph{Physics of Plasmas}, 12\penalty0 (5):\penalty0 056115, 2005.
\newblock \doi{10.1063/1.1873792}.
\newblock URL \url{http://link.aip.org/link/?PHP/12/056115/1}.

\bibitem[Boom et~al.(2011)Boom, Classen, de~Vries, Eich, Wolfrum, Suttrop,
  Wenninger, Donné, Tobias, Domier, Jr, Park, and the ASDEX
  Upgrade~Team]{Boom2011}
J.E. Boom, I.G.J. Classen, P.C. de~Vries, T.~Eich, E.~Wolfrum, W.~Suttrop, R.P.
  Wenninger, A.J.H. Donné, B.J. Tobias, C.W. Domier, N.C.~Luhmann Jr, H.K.
  Park, and the ASDEX Upgrade~Team.
\newblock {2D} {ECE} measurements of type-{I} edge localized modes at {ASDEX
  Upgrade}.
\newblock \emph{Nuclear Fusion}, 51\penalty0 (10):\penalty0 103039, 2011.
\newblock URL \url{http://stacks.iop.org/0029-5515/51/i=10/a=103039}.

\bibitem[Wenninger et~al.()Wenninger, Zohm, Boom, Eich, Garcia-Munoz, Hoelzl,
  Lunt, Maraschek, Mueller, Sommer, Viezzer, and the {ASDEX Upgrade
  Team}]{WENNINGER2011}
R~P Wenninger, H~Zohm, J~Boom, T~Eich, M~Garcia-Munoz, M~Hoelzl, T~Lunt,
  M~Maraschek, H~W Mueller, F~Sommer, E~Viezzer, and the {ASDEX Upgrade Team}.
\newblock Solitary magnetic perturbations at the {ELM} onset.
\newblock \emph{Nuclear Fusion}.
\newblock accepted.

\bibitem[Reimerdes et~al.(1998)Reimerdes, Pochelon, and Suttrop]{Reimerdes1998}
H.~Reimerdes, A.~Pochelon, and W.~Suttrop.
\newblock Toroidally asymmetric {ELM} precursors in {TCV}.
\newblock \emph{Nuclear Fusion}, 38\penalty0 (3):\penalty0 319, 1998.
\newblock URL \url{http://stacks.iop.org/0029-5515/38/i=3/a=101}.

\bibitem[B{\'e}coulet et~al.(2003)B{\'e}coulet, Huysmans, Sarazin, Garbet,
  Ghendrih, Rimini, Joffrin, Litaudon, Monier-Garbet, An{\'e}, Thomas, Grosman,
  Parail, Wilson, Lomas, deVries, Zastrow, Matthews, Lonnroth, Gerasimov,
  Sharapov, Gryaznevich, Counsell, Kirk, Valovic, Buttery, Loarte, Saibene,
  Sartori, Leonard, Snyder, Lao, Gohil, Evans, Moyer, Kamada, Chankin, Oyama,
  Hatae, Asakura, Tudisco, Giovannozzi, Crisanti, Perez, Koslowski, Eich, Sips,
  Horton, Hermann, Lang, Stober, Suttrop, Beyer, Saarelma, and {Contributors to
  JET-EFDA Workprogramme}]{Becoulet2003}
M~B{\'e}coulet, G~Huysmans, Y~Sarazin, X~Garbet, Ph~Ghendrih, F~Rimini,
  E~Joffrin, X~Litaudon, P~Monier-Garbet, J-M An{\'e}, P~Thomas, A~Grosman,
  V~Parail, H~Wilson, P~Lomas, P~deVries, K-D Zastrow, G~F Matthews,
  J~Lonnroth, S~Gerasimov, S~Sharapov, M~Gryaznevich, G~Counsell, A~Kirk,
  M~Valovic, R~Buttery, A~Loarte, G~Saibene, R~Sartori, A~Leonard, P~Snyder,
  L~L Lao, P~Gohil, T~E Evans, R~A Moyer, Y~Kamada, A~Chankin, N~Oyama,
  T~Hatae, N~Asakura, O~Tudisco, E~Giovannozzi, F~Crisanti, C~P Perez, H~R
  Koslowski, T~Eich, A~Sips, L~Horton, A~Hermann, P~Lang, J~Stober, W~Suttrop,
  P~Beyer, S~Saarelma, and {Contributors to JET-EFDA Workprogramme}.
\newblock Edge localized mode physics and operational aspects in tokamaks.
\newblock \emph{Plasma Physics and Controlled Fusion}, 45\penalty0
  (12A):\penalty0 A93, 2003.
\newblock URL \url{http://stacks.iop.org/0741-3335/45/i=12A/a=007}.

\bibitem[Eich et~al.(2003)Eich, Herrmann, and Neuhauser]{Eich2003}
T.~Eich, A.~Herrmann, and J.~Neuhauser.
\newblock Nonaxisymmetric energy deposition pattern on {ASDEX Upgrade} divertor
  target plates during type-{I} edge-localized modes.
\newblock \emph{Phys. Rev. Lett.}, 91:\penalty0 195003, Nov 2003.
\newblock \doi{10.1103/PhysRevLett.91.195003}.

\bibitem[Bobkov et~al.(2004)Bobkov, Becoulet, Blackman, Brzozowski, Challis,
  Gerasimov, Lamalle, Maraschek, Mayoral, Monakhov, Noterdaeme, Saibene,
  Walden, Wouters, {ASDEX Upgrade Team}, and contributors]{Bobkov2004}
Vl.V. Bobkov, M.~Becoulet, T.~Blackman, J.~Brzozowski, C.~Challis,
  S.~Gerasimov, P.U. Lamalle, M.~Maraschek, M.-L. Mayoral, I.~Monakhov, J.-M.
  Noterdaeme, G.~Saibene, A.~Walden, P.~Wouters, {ASDEX Upgrade Team}, and
  {JET-EFDA} contributors.
\newblock Studies of {ELM} toroidal asymmetry using {ICRF} antennas at {JET}
  and {ASDEX Upgrade}.
\newblock In \emph{Europhysics Conference Abstracts}, volume 28G, 2004.
\newblock URL \url{http://epsppd.epfl.ch/London/pdf/P1_141.pdf}.

\bibitem[Wilson and Cowley(2004)]{Wilson2004}
H.~R. Wilson and S.~C. Cowley.
\newblock Theory for explosive ideal magnetohydrodynamic instabilities in
  plasmas.
\newblock \emph{Phys. Rev. Lett.}, 92:\penalty0 175006, Apr 2004.
\newblock \doi{10.1103/PhysRevLett.92.175006}.
\newblock URL \url{http://link.aps.org/doi/10.1103/PhysRevLett.92.175006}.

\bibitem[Kirk et~al.(2005)Kirk, Eich, Herrmann, Muller, Horton, Counsell,
  Price, Rohde, Bobkov, Kurzan, Neuhauser, Wilson, the {ASDEX Upgrade}, and
  Teams]{Kirk2005}
A~Kirk, T~Eich, A~Herrmann, H~W Muller, L~D Horton, G~F Counsell, M~Price,
  V~Rohde, V~Bobkov, B~Kurzan, J~Neuhauser, H~Wilson, the {ASDEX Upgrade}, and
  {MAST} Teams.
\newblock The spatial structure of type-{I} {ELMs} at the mid-plane in {ASDEX
  Upgrade} and a comparison with data from {MAST}.
\newblock \emph{Plasma Physics and Controlled Fusion}, 47\penalty0
  (7):\penalty0 995, 2005.
\newblock URL \url{http://stacks.iop.org/0741-3335/47/i=7/a=003}.

\bibitem[Sauter et~al.(1997)Sauter, Haye, Chang, Gates, Kamada, Zohm, Bondeson,
  Boucher, Callen, Chu, Gianakon, Gruber, Harvey, Hegna, Lao, Monticello,
  Perkins, Pletzer, Reiman, Rosenbluth, Strait, Taylor, Turnbull, Waelbroeck,
  Wesley, Wilson, and Yoshino]{Sauter1997}
O.~Sauter, R.~J.~La Haye, Z.~Chang, D.~A. Gates, Y.~Kamada, H.~Zohm,
  A.~Bondeson, D.~Boucher, J.~D. Callen, M.~S. Chu, T.~A. Gianakon, O.~Gruber,
  R.~W. Harvey, C.~C. Hegna, L.~L. Lao, D.~A. Monticello, F.~Perkins,
  A.~Pletzer, A.~H. Reiman, M.~Rosenbluth, E.~J. Strait, T.~S. Taylor, A.~D.
  Turnbull, F.~Waelbroeck, J.~C. Wesley, H.~R. Wilson, and R.~Yoshino.
\newblock Beta limits in long-pulse tokamak discharges.
\newblock \emph{Physics of Plasmas}, 4\penalty0 (5):\penalty0 1654--1664, 1997.
\newblock \doi{10.1063/1.872270}.
\newblock URL \url{http://link.aip.org/link/?PHP/4/1654/1}.

\bibitem[Schmid et~al.(2008)Schmid, Herrmann, Müller, and the ASDEX
  Upgrade~Team]{Schmid2008}
A~Schmid, A~Herrmann, H~W Müller, and the ASDEX Upgrade~Team.
\newblock Experimental observation of the radial propagation of {ELM} induced
  filaments on {ASDEX Upgrade}.
\newblock \emph{Plasma Physics and Controlled Fusion}, 50\penalty0
  (4):\penalty0 045007, 2008.
\newblock URL \url{http://stacks.iop.org/0741-3335/50/i=4/a=045007}.

\bibitem[Kirk et~al.(2011)Kirk, Muller, Herrmann, Kocan, Rohde, Tamain, and the
  ASDEX Upgrade~Team]{Kirk2011}
A~Kirk, H~W Muller, A~Herrmann, M~Kocan, V~Rohde, P~Tamain, and the ASDEX
  Upgrade~Team.
\newblock Parameters determining the radial propagation of type-{I} edge
  localized modes ({ELMs}) in {ASDEX Upgrade}.
\newblock \emph{Plasma Physics and Controlled Fusion}, 53\penalty0
  (3):\penalty0 035003, 2011.
\newblock URL \url{http://stacks.iop.org/0741-3335/53/i=3/a=035003}.

\end{thebibliography}


\end{document}